\documentclass[useAMS,usenatbib]{mn2e}
\usepackage{graphicx,times}
\newcommand\HII{\mbox{H\thinspace{\sc ii}}}
\newcommand{\Msol}{$M_\odot$}
\newcommand{\coet}{\mbox{C$^{18}$O}}
\newcommand{\cotw}{\mbox{$^{12}$CO}}
\newcommand{\cott}{\mbox{$^{13}$CO}}

\newcommand{\tmb}{\mbox{$T_{\rm mb}$}}
\newcommand{\tex}{\mbox{$T_{\rm ex}$}}
\newcommand{\tas}{\mbox{$T_{\rm A}^*$}}
\newcommand{\kms}{\mbox{km~s$^{-1}$}}
\title[Mapping the column density in the GMC associated with RCW 106]{Molecular line mapping of the giant molecular cloud associated with RCW 106 Ð-- II. Column density and dynamical state of the clumps}

\author[T. Wong et al.]{T. Wong$^{1,2,3}$\thanks{Email: wongt@astro.uiuc.edu},
  E. F. Ladd$^{4}$, D. Brisbin$^{4}$, M. G. Burton$^{1}$, I. Bains$^{1,5}$, 
  M. R. Cunningham$^{1}$, 
  \newauthor
  N. Lo$^{1,3}$,  P. A. Jones$^{1}$, K. L. Thomas$^{6}$, S. N. 
  Longmore$^{1}$, A. Vigan$^{1}$, B. Mookerjea$^{7,8}$,
  \newauthor
  C. Kramer$^{7}$, Y. Fukui$^{9}$, A. Kawamura$^9$\\
  \\ $^1$ School of Physics, University of New South Wales, Sydney, NSW 2052, Australia
  \\ $^2$ Astronomy Department, University of Illinois, 1002 W. Green St,
  Urbana, IL 61801, USA
  \\ $^3$ CSIRO Australia Telescope National Facility,
  PO Box 76, Epping, NSW 1710, Australia
  \\ $^4$ Department of Physics and Astronomy, Bucknell University, Lewisburg,
  PA 17837, USA
  \\ $^5$ Centre for Astrophysics and Supercomputing, Swinburne University of Technology, PO Box 218, Hawthorn, VIC 3122, Australia
  \\ $^6$ Physics \& Astronomy Department, University of Kentucky, Lexington, KY 40506, USA
  \\ $^7$ KOSMA, I. Physikalisches Institut, Universit\"at zu K\"oln, Z\"ulpicher Stra\ss e 77, 50937 K\"oln, Germany
  \\ $^8$ Department of Astronomy \& Astrophysics, Tata Institute of Fundamental 
Research, Homi Bhabha Road, Colaba, Mumbai, India
  \\ $^9$ Department of Physics, Nagoya University, Chikusa-ku, Nagoya 464-8602, Japan}

\begin{document}

\date{Accepted 2008 February 12. Received 2008 January 24; in original form 2007 September 17}

\maketitle

\begin{abstract}\label{sec:abstract}

We present a fully sampled \coet\ (1--0) map towards the southern giant molecular cloud (GMC) associated with the \HII\ region RCW 106, and use it in combination with previous \cott\ (1--0) mapping to estimate the gas column density as a function of position and velocity.  We find localized regions of significant \cott\ optical depth in the northern part of the cloud, with several of the high-opacity clouds in this region likely associated with a limb-brightened shell around the \HII\ region G333.6$-$0.2.  Optical depth corrections broaden the distribution of column densities in the cloud, yielding a log-normal distribution as predicted by simulations of turbulence.  Decomposing the \cott\ and \coet\ data cubes into clumps, we find relatively weak correlations between size and linewidth, and a more sensitive dependence of luminosity on size than would be predicted by a constant average column density.  The clump mass spectrum has a slope near $-1.7$, consistent with previous studies.  The most massive clumps appear to have gravitational binding energies well in excess of virial equilibrium; we discuss possible explanations, which include magnetic support and neglect of time-varying surface terms in the virial theorem.  Unlike molecular clouds as a whole, the clumps within the RCW 106 GMC, while elongated, appear to show random orientations with respect to the Galactic plane.

\end{abstract}


\section{Introduction}

Molecular clouds must be supported by turbulent motions in order to resist gravitational collapse and thereby explain the relatively long time-scale for star formation in the Galaxy \citep{Zuckerman:74}.  Observations of clumpy and filamentary structures in these clouds, in qualitative agreement with numerical simulations of turbulence, provide further evidence for turbulent support.  However, the simulations also indicate that turbulence dissipates quickly (within a few crossing times, e.g. \citealt{Stone:98}), and must therefore be continually renewed by energy injection.  Given that several physical processes, acting on different scales and contributing different amounts of kinetic energy, can contribute to turbulence, the nature of the principal driving source of turbulence in molecular clouds remains a subject of active debate.  

When studying the turbulent properties of molecular clouds and their possible universality, statistics which distinguish among different spatial scales are generally preferred: these include Fourier power spectra of 2-D velocity slices \citep{Lazarian:00}, the delta variance \citep{Stutzki:98}, and principal component analysis \citep{Heyer:97}.  These techniques take advantage of the self-similar nature of turbulence to reduce its characterization to a small number of quantities, e.g.\ power-law indices of the energy spectrum or of the correlation between spatial scale and velocity dispersion.  A number of studies in recent years have applied these methods to atomic and molecular line data \citep{Dickey:01,Bensch:01,Brunt:02,Sun:06}.

On the other hand, relating turbulence to star formation requires that the molecular gas data be considered as a function of position, and not only as a function of spatial scale.  To facilitate such analysis, it is commonplace to decompose molecular line maps into discrete structures, usually referred to as clumps or cores, and to relate the properties of these clumps and cores to the properties of stars and star-forming units.  Although several automated decomposition methods have been introduced (e.g., GAUSSCLUMPS, \citealt{Stutzki:90}; CLUMPFIND, \citealt{Williams:94}) all are sensitive to the map resolution and to {\it ad hoc} parameters selected by the user.  Papers emphasizing numerical results have tended to cast doubt on the interpretation of these decompositions \citep{Ballesteros:02,Dib:07}, yet they remain useful as a way to identify regions that are amenable to multi-wavelength analysis.

Motivated by an interest in both approaches to studying molecular cloud structure, we have recently undertaken a systematic program of observing the giant molecular cloud (GMC) associated with the RCW 106 star-forming region in multiple molecular lines with the Mopra Telescope.  \citet[][hereafter Paper I]{Bains:06} presented the initial results of the program.  Applying a two-dimensional clump-finding algorithm to an integrated \cott\ ($J$=1 $\rightarrow$ 0) emission map, we found that about half of the identified \cott\ clumps could not be distinguished as clumps in the 1.2-mm dust emission map made by \citet{Mookerjea:04}.  While some of these clumps may have fallen below the sensitivity limit of the dust map, it is possible that a significant fraction of the gas clumps were the result of integrating unrelated cloud structures in velocity.

In this paper we present our \coet\ ($J$=1 $\rightarrow$ 0) mapping of this region, and derive improved estimates of molecular column density across the cloud.  We also quantify and discuss variations in \cott\ optical depth and their implications for estimating the probability density function (PDF) of the column density.  Finally, we revisit the structural decomposition of the cloud emission, working directly on the three-dimensional data cubes, using the CPROPS algorithm developed by \citet[][hereafter RL06]{Rosolowsky:06}.  Future papers in this series will examine the power spectra of the molecular gas (P. Jones {et~al.}, in preparation) and the relationship between recent star formation and the molecular clumps (N. Lo {et~al.}, in preparation).  As in Paper I, we adopt a distance to the cloud of 3.6 kpc \citep{Lockman:79}, giving it a projected size of roughly 90 $\times$ 30 pc.  Since all observed lines are $J$=1 $\rightarrow$ 0 transitions, we simply refer to them hereafter by the name of the species (`\cott', etc.).

\section{Observations}

\begin{figure}
\includegraphics[width=8.5cm]{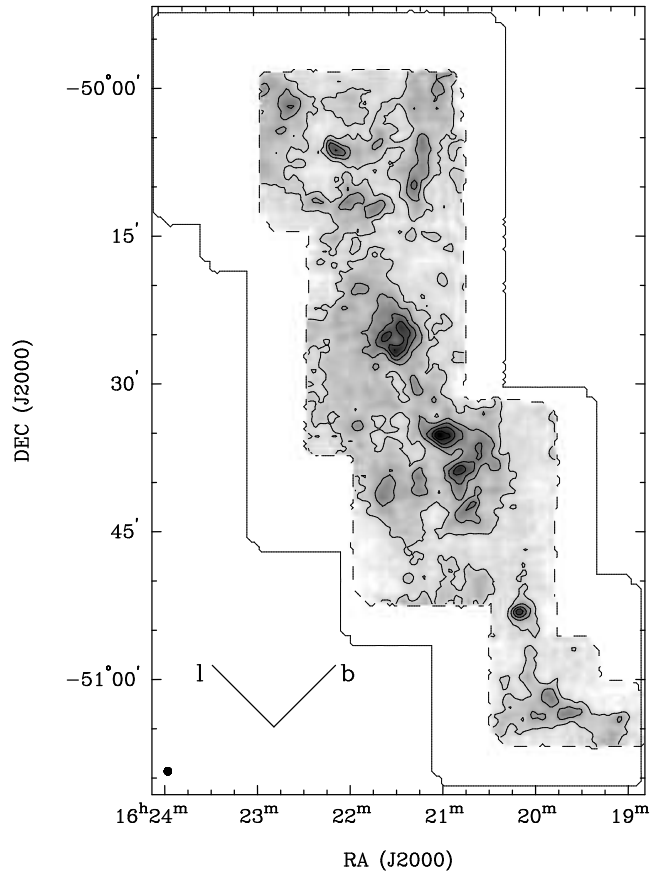}
\caption{\coet\ map integrated from $-65$ to $-35$ \kms.  Contour levels are linearly spaced starting at 7 K \kms\ (\tmb).  The dashed line represents the region mapped with Mopra, while the larger solid line represents the region previously mapped in \cott\ (Paper I).  The Mopra beam size is shown in the lower left corner.  The orientation of the Galactic coordinate axes is also shown at lower left.}
\label{fig:dqsc18o}
\end{figure}

Observations of the \coet\ ($J$=1 $\rightarrow$ 0) line towards RCW 106 were made with the 22-m Mopra telescope of the Australia Telescope National Facility (ATNF) between 2005 July 5 and September 11.  A total of 46 5\arcmin\ $\times$ 5\arcmin\ fields were observed in two passes, with the first pass produced by scanning along the R.A. axis and the second pass along the Dec.\ axis.  The fields observed in each pass were chosen to have some overlap, with centre positions separated by 4\farcm 75, as for the \cott\ observations described in Paper I.  In addition, the field centres for the second pass were shifted by 1\arcmin\ in both R.A. and Dec.\ (towards the southwest) in order to further reduce sensitivity variations at the field edges.  The region mapped is shown in Figure~\ref{fig:dqsc18o}, and is a subset of the region previously mapped in \cott, which comprised 93 fields.

The observations were conducted in raster scan (`on-the-fly') mode, with the same mapping parameters and speed as for \cott\ (Paper I).  Each field took $\sim$70 min to complete, and the system temperature was calibrated every $\sim$30 min using an ambient load.  Typical above-atmosphere system temperatures were $\sim$300 K.  The pointing solution was refined after each field using observations of the nearby SiO maser IRSV 1540; typical corrections were $\sim$5\arcsec.  

The Mopra digital correlator was configured to output 1024 channels across a 64 MHz bandwidth, providing a velocity resolution of 0.17 \kms.  To ensure that reference and source observations were taken at the same frequency, Doppler tracking was disabled and the center frequency was set to 109.792 GHz, which corresponded to $v_{\rm LSR} \approx -50$ \kms\ at the time of the observations.  As in Paper I, we quote velocities in this paper relative to the local standard of rest (LSR).

Data reduction was performed using the ATNF {\sc MIRIAD}, {\sc LIVEDATA} and {\sc GRIDZILLA} packages.\footnote{http://www.atnf.csiro.au/computing/software/}  Following data inspection and editing, spectra were bandpass corrected using an emission-free spectrum taken at the beginning of each row at a fixed sky position ($\alpha_{2000}$=16:27, $\delta_{2000}$=$-51$:30).  A linear baseline was fitted to channels deemed to be emission-free and subtracted.  The spectra were then gridded into a datacube using a 33\arcsec\ (FWHM) Gaussian smoothing kernel and 15\arcsec\ pixel scale.  Approximating the Mopra point source response as a Gaussian with FWHM 33\arcsec, the final map resolution was determined to be 45\arcsec.  The RMS noise in the map was 0.12$\pm$0.07 K (\tas) and ranged from 0.063 to 1~K; high-noise regions at the edge of the map were masked out by setting a threshold RMS of 0.25~K.\@  The resulting masked map has an RMS noise of 0.11$\pm$0.03 K.  Figure~\ref{fig:dqsc18o} shows the \coet\ intensity integrated from $-65$ to $-35$ \kms, a velocity range which covers the bulk of the emission.

Since the Mopra beam response is not perfectly Gaussian, a correction factor is needed to account for flux which is scattered out of the main lobe of the primary beam.  \citet{Ladd:05} derived a main-beam efficiency for the 2004 season of $\eta_{mb}$=0.42 at 115 GHz, and an `extended' beam efficiency (for sources that couple to the inner error beam) of $\eta_{xb}$=0.55.  Because the structures of interest for this paper are comparable to or larger than the size of the inner error beam ($\sim$80\arcsec), the extended beam efficiency is used here.  To determine whether there was a change in $\eta_{xb}$ from 2004 to 2005, we compared uncorrected (\tas\ scale) spectra taken by one of us (BM) in 2004 June towards 15 positions in the RCW 106 cloud with the 2005 map data (with minimal smoothing applied in the gridding process).  For 449 channels with \tas$>$1~K, the average ratio of 2005 to 2004 intensities was 0.9 with a standard deviation of 0.18.  Since the data suggest a change of $\sim$10 per cent in the efficiency, we adopt a value of $\eta_{xb}$=0.5 to correct the intensity scale to \tmb\ units, although the intensity calibration is likely to remain uncertain at the $\sim$10--20 per cent level.

\begin{figure}
\includegraphics[width=8.5cm,bb=39 307 549 675]{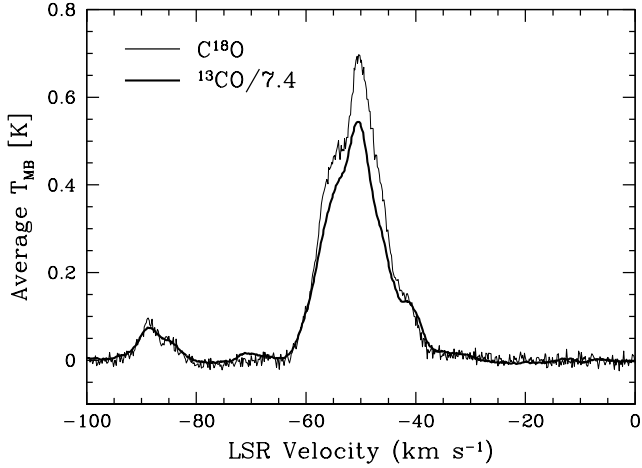}
\caption{Average \coet\ spectrum, in \tmb\ units, over the region shown in Fig.~\ref{fig:dqsc18o}.  The thick line is the \cott\ spectrum averaged over the same region, divided by the adopted abundance ratio of 7.4.}
\label{fig:intspec}
\end{figure}

Figure~\ref{fig:intspec} shows a spectrum of the \coet\ brightness temperature averaged over the map.  The \cott\ brightness temperature over the same region is also shown, scaled down by the assumed abundance ratio of 7.4 (see \S\ref{sec:tau}).  Aside from calibration errors, any mismatch between the profiles could be due to variations in the abundance ratio and/or to optical depth effects.

\section{Analysis}

\begin{figure*}
\includegraphics[width=12cm,bb=100 15 515 780]{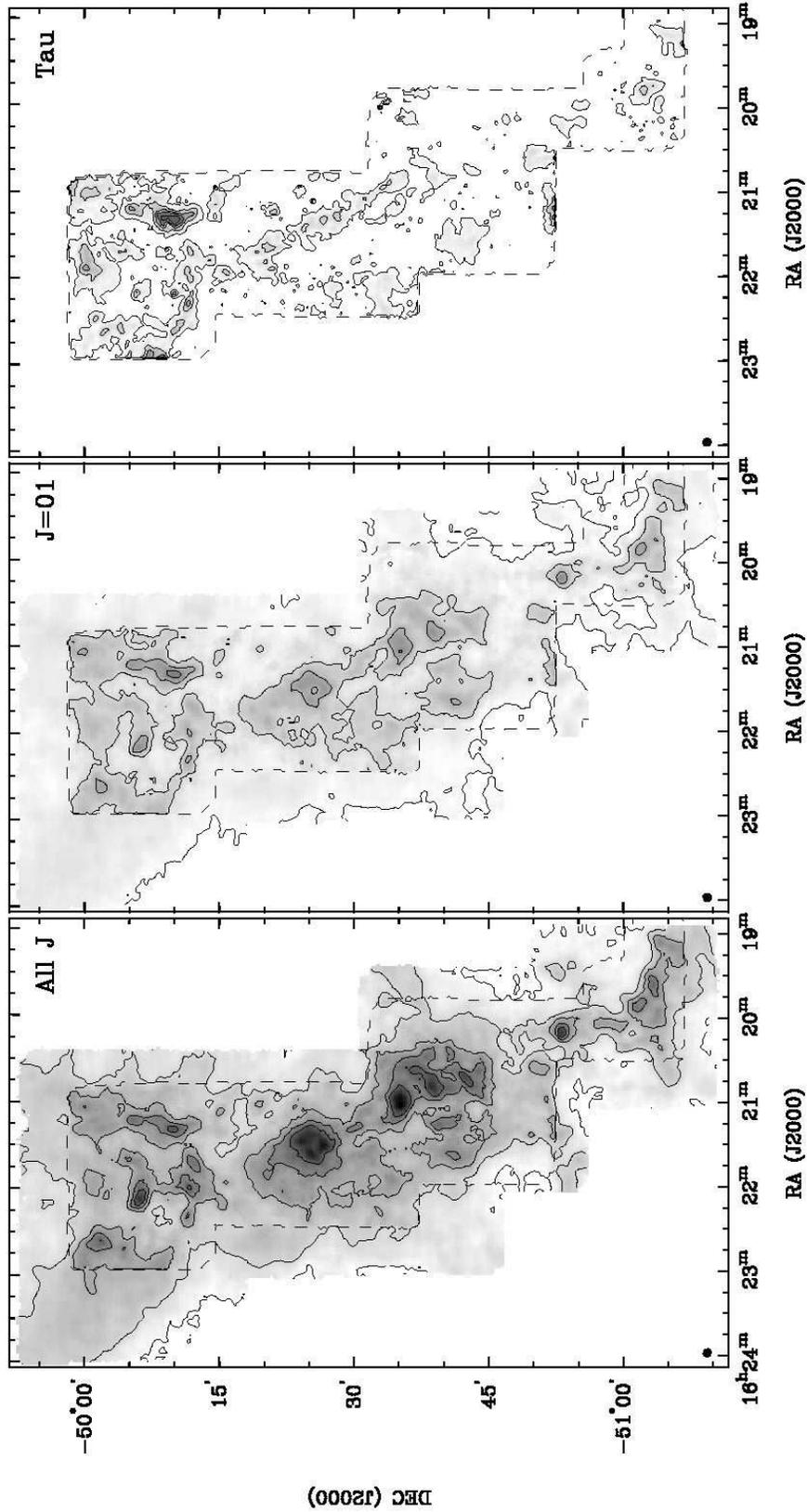}
\caption{Images of integrated column density and peak opacity.  {\it Left:} Integrated H$_2$ column density assuming all rotational levels are thermalized.  Contour levels are $6.4n^2 \times 10^{21}$ cm$^{-2}$ ($n$=1,2,3,4,5).  This gives an upper limit to the column density.  {\it Centre:} Integrated H$_2$ column density assuming only the $J$=0 and 1 states are populated.  Contour levels are again $6.4n^2 \times 10^{21}$ cm$^{-2}$.  This represents a lower limit to the column density.  {\it Right:} Peak \cott\ optical depth at each position in the map.  Contour levels are 2.4$n$ ($n$=1,2,3,4,5).}
\label{fig:colden}
\end{figure*}

\subsection{Opacity, excitation temperature and column density}\label{sec:tau}

Column densities were derived from the \cott\ and \coet\ data by assuming a filled beam and a uniform excitation temperature, common to both tracers, within the beam.  The first step was to derive the opacity in both lines at each ($\alpha$, $\delta$, $v$) volume pixel (voxel) in the datacube from the ratio of \cott\ to \coet\ main-beam brightness temperatures \citep[e.g.,][]{Myers:83}:
\begin{equation}
\frac{\tmb(\cott)}{\tmb(\coet)} = \frac{1-\exp(-\tau_{13})}{1-\exp(-\tau_{18})} = \frac{1-\exp(-7.4\tau_{18})}{1-\exp(-\tau_{18})}\;.
\label{eqn:tau18}
\end{equation}
To reduce the effects of noise, the spectra were binned to a velocity resolution of 0.68 \kms, and the brightness ratio was only computed when the signal-to-noise ratio (SNR) was $>$5 in the \coet\ data.  Voxels which failed this test, or had brightness ratios $<$1 (all such voxels were found to be consistent with noise), were assigned \cott\ column densities in the optically thin limit.

Equation~\ref{eqn:tau18} assumes $\tau_{13} = 7.4 \tau_{18}$, based on the abundance ratio [$^{13}$C][$^{16}$O]/[$^{12}$C][$^{18}$O]=7.4 estimated at a galactocentric radius of $D_{\rm GC}$=5.5 kpc \citep{Wilson:94}.  Note that this value is $\sim$30 per cent higher than the value of [$^{13}$CO]/[C$^{18}$O]=5.5 used by \citet{Myers:83}, which assumes terrestrial abundances.  Our estimate, while still quite uncertain, takes into consideration the isotopic abundance gradients in the Galaxy and the decrease in [$^{12}$C]/[$^{13}$C] since the epoch of the Earth's formation due to CNO processing.

The excitation temperature \tex\ can then be derived from the equation of radiative transfer,
\begin{equation}
\tmb = f[J(\tex)-J(T_{bg})][1-\exp(-\tau)]\;,
\label{eqn:tmbdef}
\end{equation}
where $f$ is the beam filling factor, $J(T) \equiv T_0/[\exp(T_0/T)-1]$, $T_{bg}$=2.7 K, and $T_0 = h\nu/k$ = 5.29 K for the 1--0 transition of \cott.  This equation was applied to the \cott\ rather than \coet\ data because of the higher signal-to-noise of the former.  For most of the analysis we assume a filled beam, so $f$=1.  However, in cases where the emission was weak but $\tau$ was found to be significant, resulting in \tex$<$5 K, we set \tex=5 K and calculated the filling factor $f$ needed to account for the observed brightness temperature.  Such regions of very low implied \tex\ tend to be found at the edges of the cloud (in position and velocity), where incomplete beam filling is quite likely.  We also imposed a maximum $\tau_{13}$ of 12, corresponding to a brightness temperature ratio of 1.25, as we cannot accurately measure higher opacities given uncertainties in telescope calibration and the abundance ratio.  Such voxels were extremely rare ($<$0.02 per cent) and only found in the highest opacity region in the northwest part of the map.

Given \tex, $\tau_{13}$, and $\tau_{18}$, the column densities of \cott\ and \coet\ can be derived using the expression for column density in a single rotational level \citep[e.g.,][]{Garden:91}:
\begin{equation}
N_J = \frac{3h}{8\pi^3\mu^2}\;\frac{2J+1}{J+1}\,\left[1-\exp\left(-\frac{h\nu}{k\tex}\right)\right]^{-1} \int \tau_v\; dv \;,
\end{equation}
where $J$ is the lower state of the transition and $\mu$ is the permanent dipole moment of the molecule ($\mu$=0.11 for the lines studied here).  A lower limit to the total column density $N_{\rm tot}$ can be derived by assuming only the first two levels ($J$=0 and 1) are populated \citep{Frerking:82}, while an upper limit can be derived assuming all rotational levels are thermalized with the same excitation temperature \tex.  The latter assumption yields \citep{Garden:91}:
\begin{equation}
N_{\rm tot} = \frac{3k}{8\pi^3B\mu^2}\; \frac{\tex + hB/3k}{1-\exp(-h\nu/k\tex)}\; \int \tau_v\; dv\;,
\label{eqn:ntot}
\end{equation}
where $B$ is the rotational constant.

In a significant fraction of voxels (roughly 12 per cent of the total), generally with detected but weak \coet\ emission, we found $\tmb(\cott)/\tmb(\coet) > 7.4$, resulting in optical depths which were undefined in Equation~\ref{eqn:tau18}.  The largest significant values for the ratio are $\approx$14 and occur near ($\alpha$, $\delta$, $v$) = (16:20:53, $-50$:43:55, $-56$).  These high values could result from a slight mismatch in beam sizes near regions of bright emission, from differences in excitation between \cott\ and \coet, or from chemical fractionation increasing the abundance of \cott\ relative to \cotw\ and \coet\ \citep{Langer:80}.  For these cases we assumed an excitation temperature based on the \cotw\ brightness temperature as described in \S\ref{sec:nanten}, and derived $N(\coet)$ in the optically thin limit, where
\[T_{\rm mb} \approx [J(\tex)-J(T_{bg})]\tau \;.\]
This yields \citep[cf.][]{Bourke:97}:
\begin{eqnarray}
\lefteqn{N(\coet)_{\rm thin} = 2.42 \times 10^{14}\; 
\frac{\tex+0.88}{1-\exp(-5.27/\tex)}\;\times} \hspace{2cm} \nonumber\\
& & \frac{1}{J(\tex)-J(T_{bg})}\; \int T_{\rm mb}(\coet)\; dv\; .
\end{eqnarray}
We also applied this procedure whenever $\tau_{13}$$<$0.1, to limit the number of poorly constrained, very high values of \tex.  If the column densities had been calculated from \cott\ instead, they would have been on average $\sim$10 per cent higher (assuming an abundance ratio of 7.4), though in some regions up to a factor of 2 higher.

In other voxels, only \cott\ was observed or detected.  Again, an excitation temperature was assumed based on \cotw, and $N(\cott)$ was derived in the optically thin limit \citep{Bourke:97}:
\begin{eqnarray}
\lefteqn{N(\cott)_{\rm thin} = 2.42 \times 10^{14}\; 
\frac{\tex+0.88}{1-\exp(-5.29/\tex)}\;\times} \hspace{2cm} \nonumber\\
& & \frac{1}{J(\tex)-J(T_{bg})}\; \int T_{\rm mb}(\cott)\; dv\; .
\end{eqnarray}

The molecular hydrogen column density, $N(\rm H_2)$, was then calculated by integrating the \cott\ or \coet\ column density cube in velocity from $-80$ and $-20$ \kms assuming an [H$_2$/\cott] abundance ratio of $7 \times 10^5$ \citep{Frerking:82} and an [H$_2$/\coet] abundance ratio of $5 \times 10^6$ (for consistency with our adopted [\cott/\coet] abundance ratio).  Figure~\ref{fig:colden} shows the resulting column density maps (assuming thermalisation of all $J$ levels or only population of $J$=0 and 1) as well as a map showing the peak optical depth of \cott\ at each spatial position.  Note that the column density outside the region mapped in \coet\ is more uncertain due to our lack of constraints on the opacity.

\subsection{Default \tex\ using NANTEN data}\label{sec:nanten}

In order to provide an independent estimate of the excitation temperature in cases where it could not be derived from the \cott\ and \coet\ data directly, we made use of the \cotw\ data from the NANTEN Galactic Plane Survey.  The observations and data reduction have been described by \citet{Matsunaga:01}.  In brief, spectra taken with a grid spacing of 4\arcmin\ using the 4-m NANTEN telescope (half-power beamwidth 2\farcm6).  An acousto-optical spectrometer (AOS) provided velocity coverage of 650 \kms\ at a resolution of 0.65 \kms.  The resulting map was interpolated onto the Mopra grid and \tex\ for each voxel was derived from the peak $T_{\rm mb}$ at the corresponding position in the \cotw\ data cube, assuming a filled beam and optically thick CO ($\tau_{12} \gg 1$).  Where the peak \tmb(\cotw) was less than 5~K, the minimum \tex=5 K was adopted as above.  No attempt was made to deconvolve the \cotw\ map to approximate the resolution of the Mopra data, so this method is likely to yield an underestimate of \tex\ in regions of bright emission while an overestimate in adjacent regions.  In addition, the assumption of optically thick \cotw\ in a filled beam is likely to yield too low an estimate of \tex, especially at the edges of the cloud, although our imposition of a minimum \tex\ should help to reduce errors introduced by this assumption.  

\begin{figure*}
\includegraphics[width=12cm,bb=100 15 515 780]{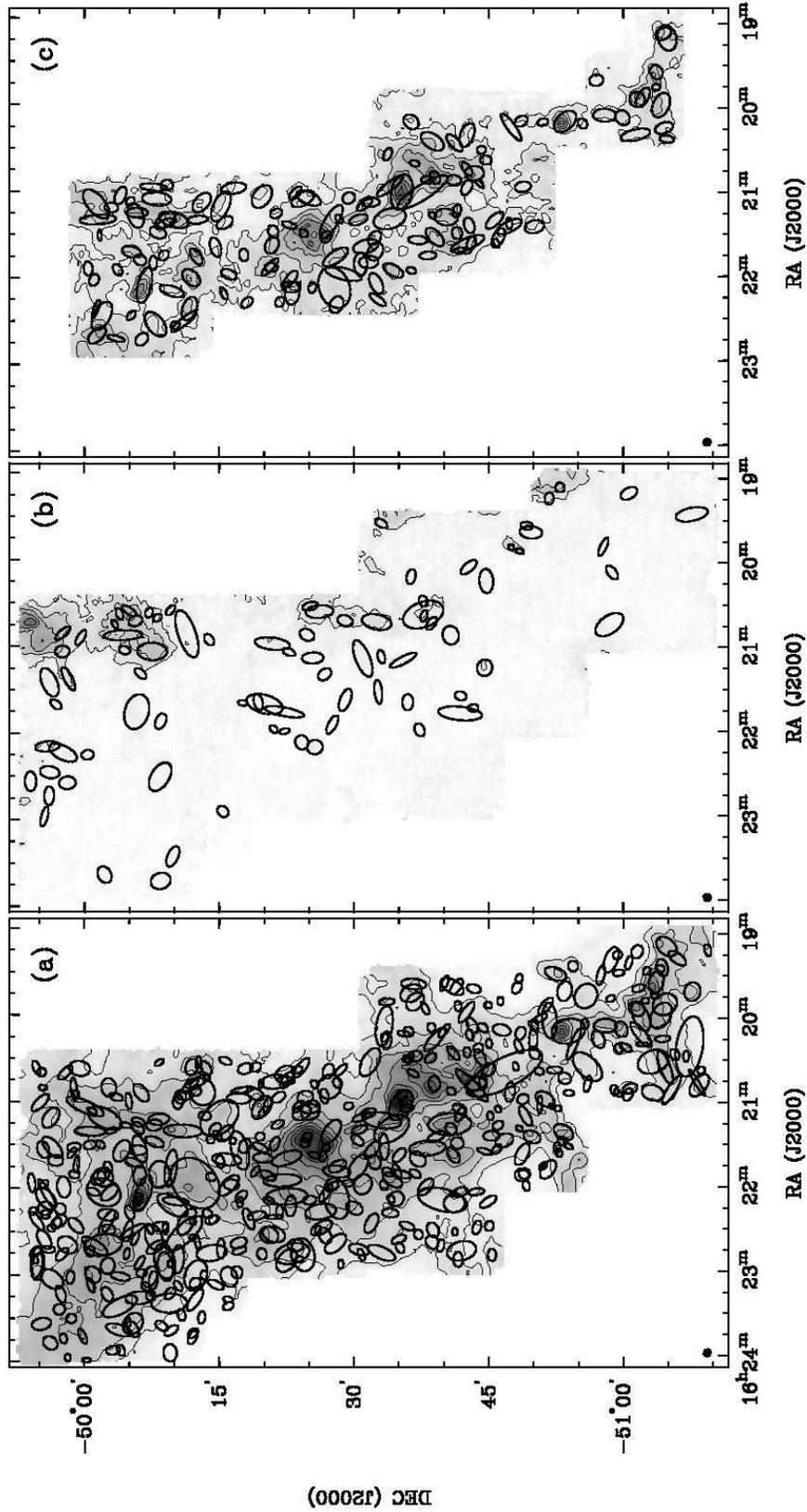}
\caption{Locations and sizes of the clumps derived from the CPROPS analysis.  Sensitivity and resolution corrections have not been applied.  The ellipse major axis represents the clump major axis as derived from principal component analysis.  (a) Integrated \cott\ emission from the GMC.  Contour levels are spaced by 10 K \kms\ (\tas). (b) Non-GMC integrated \cott\ emission.  Contour levels are spaced by 3 K \kms\ (\tas). (c) Integrated \coet\ emission.  Contour levels are spaced by 3 K \kms\ (\tas).}
\label{fig:clumps}
\end{figure*}

\subsection{Decomposition into clumps}\label{sec:propanal}

Paper I had noted several velocity components in the \cott\ cube, with most of the emission centered on an LSR velocity near $-50$ \kms\ (Fig.~\ref{fig:intspec}).  In order to distinguish the main GMC emission from unrelated emission along the line of sight, we applied the `dilated mask' method in the IDL-based CPROPS package of RL06.  The \cott\ spectra were binned to 0.34 \kms\ resolution to increase the signal-to-noise ratio, and a mask was created by starting from a threshold contour of $T_A^* = 5$ K and expanding in both position and velocity to include all simply connected regions above a brightness threshold of 0.26~K (3$\sigma$).  The resulting ``GMC mask'' was applied to the \cott\ cube, yielding a single continuous region encompassing 92 per cent of the emission between $-80$ and $-20$ \kms\ and extending in velocity from $-68$ to $-25$ \kms.  Of course, such a procedure does not guarantee that unrelated emission is not projected into the same velocity range.  However, given that the region of interest is located somewhat away from the Galactic plane (from $b$=$-$0\fdg2 to $-$0\fdg6), it seems reasonable to assume that the emission within the mask comes from a single cloud.

To decompose the GMC into discrete structures, we applied a modified version of the CPROPS algorithm to the masked data cube.  Significant peaks (local maxima) in the emission were identified as voxels which are greater than all their neighbours within an $l \times l \times k$ box, where $l$=5\arcmin\ and $k$=3 \kms.  The adopted values of $l$ and $k$ were chosen to reduce blending between clumps and ensure that the decomposition could be run in a reasonable amount of time.  However, they may bias our analysis against detection of closely spaced clumps.  The data were then contoured below each peak until the highest contour level which begins to enclose another peak; this level (which is unique to each peak) is termed the {\it merge level}.  Peaks which did not lie at least 2$\sigma$ above the merge level, or did not contain at least 2 beam areas of pixels above the merge level, were discarded.  The resulting `clumps' were defined as the regions uniquely associated with each peak (i.e., above the merge level).  Unlike the original CPROPS procedure of RL06, we did not filter the list of clumps based on the change in moments when merging with adjacent clumps.  Such filtering, which has the effect of merging clumps which blend smoothly into each other, drastically reduces the number of clumps (from $\sim$600 to $\sim$30), and in our case leaves a great deal of substructure within each clump. 

The clumps derived from CPROPS contain a relatively small fraction ($\sim$10 per cent) of the total \cott\ emission, since contour levels which enclose several peaks (blended contours) are not decomposed.  As an alternative, we also performed decompositions using the {\tt /ECLUMP} option in CPROPS, which applies the CLUMPFIND algorithm of \citet{Williams:94} to the previously identified emission peaks.  In this case, blended contours, rather than being discarded, are assigned to existing clumps via a `friends-of-friends' algorithm.  This has the effect of assigning all of the flux within the GMC mask into clumps, and thus producing a `patchwork' of much larger clumps.  Unlike CLUMPFIND, however, the number of clumps cannot increase as one works down to lower contour levels: the initial number of peaks remains the final number of clumps.

We also ran both procedures discussed above on two additional cubes: a non-GMC \cott\ cube, which consists of the emission between $-80$ and $-20$ \kms\ that lies outside the GMC mask, and the whole \coet\ cube within the same velocity range.  In both cubes, islands of discrete emission were identified by starting from a threshold of 4$\sigma$ in two adjacent channels and expanding to a 3$\sigma$ boundary (still requiring significant emission in at least 2 adjacent channels).  Each island was then decomposed into clumps as was done for the GMC \cott\ emission.  Although the \coet\ cube was decomposed without the use of the GMC mask, it turned out that all but two of the resulting clumps were within the mask boundary, so effectively all of the decomposed \coet\ emission is associated with the GMC.\@  The non-GMC \cott\ cube, on the other hand, contains emission which appears separated from the GMC in position or velocity, and thus may be foreground or background emission.

The final step in the CPROPS analysis is to determine basic properties (radius, linewidth, and luminosity) for each clump.  The radius of a clump $R$ is proportional to $\sigma_r$, the geometric mean of the RMS sizes (second spatial moments) of the clump along the major and minor axes, with the major axis determined via principal component analysis.  Following RL06, we take $R=1.91 \sigma_r$ (as previously adopted by \citealt{Solomon:87}).  The FWHM linewidth $\Delta V$ is equal to 2.35$\sigma_v$, where $\sigma_v$ is the intensity-weighted second moment along the velocity axis.  The luminosity is based on the integrated flux of the cloud and the adopted distance (3.6 kpc, except for the non-GMC clumps where the near kinematic distance was used).  RL06 detail procedures to correct for the sensitivity and resolution biases that affect real observational data; this involves extrapolating the major and minor axes, linewidth, and flux to infinite sensitivity based on their values at various signal-to-noise levels, and then (for the radius and linewidth) subtracting the beam or channel width in quadrature.  In the end we did not apply these procedures, for reasons discussed further in \S\ref{sec:cprops_corr}.  The locations and sizes of the clumps are shown overlaid on the respective integrated intensity maps in Figure~\ref{fig:clumps}.

\begin{figure}
\includegraphics[width=8.5cm,bb=33 171 554 680]{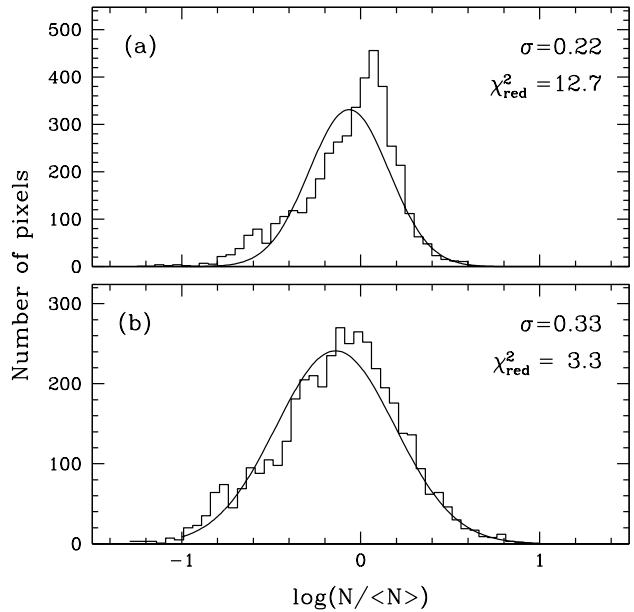}
\caption{Histogram of normalised column densities in the RCW106 GMC, integrated from $-80$ to $-20$ \kms\ over the region observed in \cott.  (a) Column density based on the optically thin assumption for \cott\ and adopting \tex=20 K.  (b) Column density calculated based on the optical depth analysis.  In regions where only \cott\ is observed, $N$ is calculated in the optically thin limit using \tex\ based on the NANTEN data.  For both panels, a best-fit Gaussian has been overplotted which represents a log-normal distribution with the dispersion indicated.}
\label{fig:nhist}
\end{figure}

\section{Results}

\subsection{Overall cloud properties}

\begin{table}
\caption{Estimates of cloud mass.}
\label{tbl:mass}
\begin{tabular}{@{}lccc}
\hline
& \cott\ map region & \coet\ map region\\
\hline
Optically thin \cott, all $J$ & 2.0 & 1.3\\
Optically thin \coet, all $J$ & --- & 1.5\\
Opacity corrected \cott, $J$=0,1 & 0.94 & 0.69\\
Opacity corrected \cott, all $J$ & 2.2 & 1.6\\
\hline
\end{tabular}
\medskip
All masses are given in units of $10^6$ \Msol.
\end{table}

The total gas mass can be estimated from the \cott\ and \coet\ cubes separately by assuming optically thin emission and a single excitation temperature for all rotational levels.  We integrated all observed emission from $-80$ to $-20$ \kms\ using the abundance ratios given at the end of \S\ref{sec:tau}.  Adopting \tex=20 K as in Paper I, and including a factor of 1.36 for helium, the resulting masses are shown in the first two lines of Table~\ref{tbl:mass}.  Note that the \cott\ estimate is corrected to be a factor of $\approx$2 higher than in Paper I, due to an error in the original calculation.  Although use of \cott\ alone may overestimate the column density in some optically thin regions, perhaps due to fractionation (\S\ref{sec:tau}), it underestimates the column density in optically thick regions, resulting in an overall mass estimate lower than that from \coet.  

The last two lines of Table~\ref{tbl:mass} lists the opacity-corrected masses derived by integration of the column density cube.  The opacity-corrected masses, assuming population of all $J$ levels, are only slightly larger than the masses derived in the optically thin approximation, which may reflect both optical depth effects and variations in \tex\ that reduce the gas mass in cold regions and increase it in warm regions.

The distribution of integrated column density values over the region mapped in \cott, after binning to 45\arcsec\ pixels to approximate the spatial resolution of the data, is shown in Figure~\ref{fig:nhist}.  Column densities $< 2 \times 10^{21}$ cm$^{-2}$ have been excluded, reflecting a 4$\sigma$ sensitivity limit of 3 K \kms\ (\tmb) in the \cott\ data.  However, imposing this cutoff has little effect, as $<$0.5 per cent of the pixels fall below this limit.  To facilitate comparison with theoretical models \citep[e.g.,][]{Vazquez:01}, we have normalized $N$ by its mean value and plotted it on a logarithmic scale.  The optically-thin, \cott-based column densities ({\it upper panel}) show a distinctly asymmetric distribution with a steep decline at high column densities.  The opacity-corrected values ({\it lower panel}) show a less abrupt decline towards high column densities, and the overall distribution can be well described as a log-normal function with a dispersion of 0.33 dex.  Underscoring this point, the best-fit Gaussian shows a significant improvement in reduced $\chi^2$ (assuming each bin has an uncertainty of $\sqrt{N}$), from 12.7 to 3.3.  Thus, although the opacity correction has little effect on the total mass estimate, it has a pronounced effect on the distribution of column densities, bringing them closer to a log-normal distribution.

\subsection{Clump orientations and masses}\label{sec:cprops_mass}

\begin{figure}
\includegraphics[width=8.5cm]{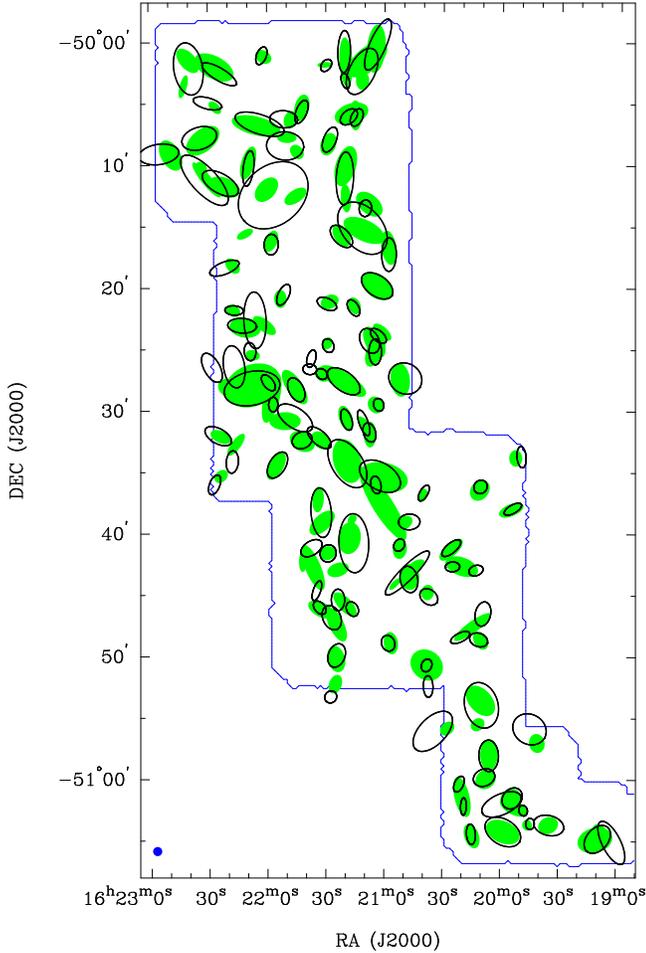}
\caption{Locations, sizes, and orientations of \cott\ clumps that overlap with \coet\ clumps in both position and velocity (open ellipses) and \coet\ clumps that overlap with \cott\ clumps (shaded ellipses).  The outer contour indicates the region mapped in \coet.}
\label{fig:olap}
\end{figure}

\begin{figure}
\includegraphics[width=8.5cm,bb=30 164 550 680]{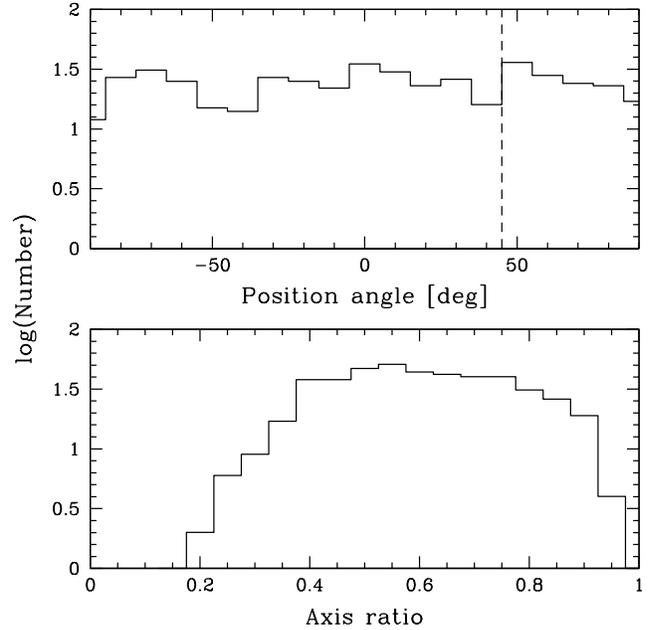}
\caption{Distribution of position angles (measured from north towards east on the sky) and minor-to-major axis ratios ($b/a$) for the \cott\ clumps.  The Galactic plane has a position angle of approximately 45\degr\ (vertical dashed line).}
\label{fig:pahist}
\end{figure}

\begin{figure}
\includegraphics[width=8.5cm,bb=30 155 554 680]{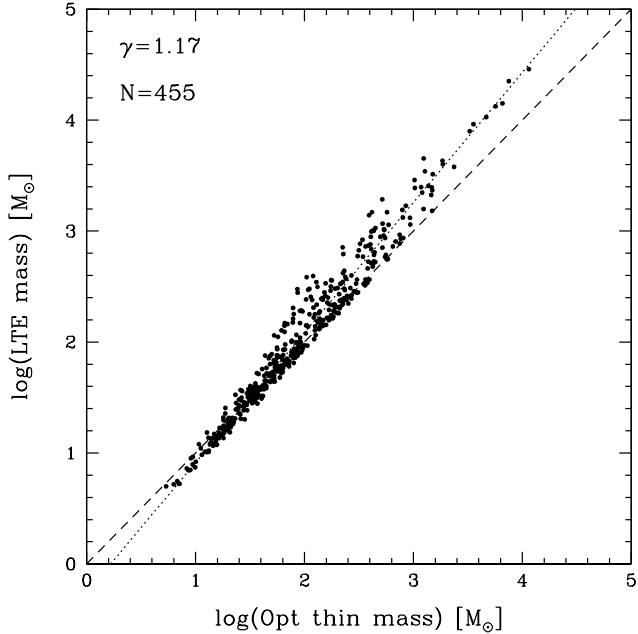}
\caption{Comparison between optically thin \cott\ mass, based on \tex=20 K, and opacity-corrected LTE mass calculated in this paper.  Note that a factor of 2 corresponds to 0.3 dex.  The dotted line is the ordinary least-squares bisector \citep[as defined by][]{Feigelson:92}, whereas the dashed line represents a slope of 1.}
\label{fig:mlte}
\end{figure}

\begin{table}
\caption{Number of clumps found with CPROPS.}
\label{tbl:cprops}
\begin{tabular}{@{}lccc}
\hline
& \cott, GMC & \cott, non-GMC & \coet, GMC\\
\hline
Initial number & 594 & 121 & 151\\
After edge culling & 456 (345) & 83 (71) & 138 (116)\\
After all culling & 455 (345) & 83 (71) & 135 (114)\\
\hline
\end{tabular}
\medskip
Numbers in parentheses correspond to the ECLUMP method.
\end{table}

As noted in \S\ref{sec:propanal}, we decomposed the cloud into clumps using two methods: the `default' method in CPROPS and the `ECLUMP' method which applies a modified CLUMPFIND algorithm.  Although the initial number of clumps from the two methods was the same, we then proceeded to exclude clumps which extended to the edge of the observed field, as their sizes and fluxes could not be reliably determined.  Because the ECLUMP method produces larger clumps, it yielded more clumps which had to be excluded.  Additional culling was then applied to remove clumps with undefined linewidths or radii, or which lay outside the GMC boundary in the case of the \coet\ cube.  As summarized in Table~\ref{tbl:cprops}, the final numbers of clumps were 455, 83, and 135 in the GMC-only \cott\ cube, non-GMC \cott\ cube, and \coet\ cube respectively.  We find a large fraction ($\sim$75\%) of the \coet\ clumps overlap in position-velocity space with \cott\ clumps, although the reverse is not true, as a result of the smaller region observed in \coet\ and the higher signal-to-noise of the \cott\ observations.  The locations, sizes, and orientations of the 223 overlapping \cott\ and \coet\ clumps are shown in Figure~\ref{fig:olap}, indicating that when there is overlap one typically finds good agreement in terms of clump size and orientation.

Figure~\ref{fig:pahist} shows the distributions of position angles and axis ratios for the GMC-only \cott\ clumps, as derived using the default CPROPS method.  In general the clumps show significant ellipticity, with a mean axis ratio $\approx$0.6, similar to the mean value found by \citet{Myers:91} for cores within dark clouds.  Since the axis lengths have not been corrected for beam smearing, the true ellipticity could be even greater.  Given the filamentary appearance of the GMC, this is not unexpected.  However the position angles of the clumps appear to be randomly oriented, with no clear preference to be aligned with the Galactic plane.  We discuss this observation further in \S\ref{sec:padisc}.

CPROPS provides two estimates of clump mass, a `luminous mass' $M_{\rm LUM}$ which is simply proportional to the observed line flux, and a `virial mass' $M_{\rm VIR}$ defined as the mass of a spherical cloud with a density profile $\rho \propto r^{-1}$ in equilibrium between internal velocity dispersion and gravity \citep{Solomon:87}:
\begin{eqnarray*}
M_{\rm VIR} & = & 189\; M_\odot\; \Delta V^2 R\\
& = & 1050\; M_\odot\; \sigma_v^2 R\,,
\end{eqnarray*}
where $\sigma_v$ is the velocity dispersion in \kms\ and $R$ is the effective radius in pc ($R=1.91 \sigma_r$).  We also calculated a `local thermodynamic equilibrium mass' $M_{\rm LTE}$ for each clump, being the mass in the corresponding voxels of the H$_2$ column density cube assuming a thermalised population of $J$-levels.  As Figure~\ref{fig:mlte} indicates, the relation between $M_{\rm LTE}$ and $M_{\rm LUM}$ is roughly linear, with only a slight non-linearity appearing at larger masses, where the luminous mass (essentially an optically thin estimate) underestimates the LTE mass by a factor of $\sim$2.5.  The virial mass, on the other hand, is likely to be a poor estimator of the mass as it relies on the assumption of virial equilibrium (\S\ref{sec:avirdisc}). 

\begin{figure}
\includegraphics[width=8.5cm,bb=40 265 550 640]{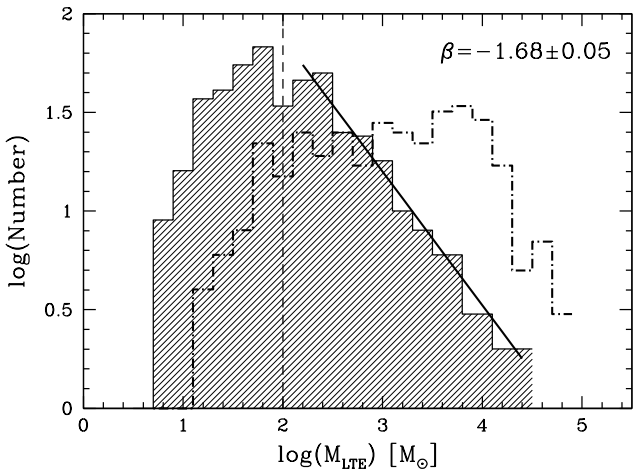}
\caption{Distribution of opacity-corrected masses for the clumps.  The shaded histogram shows results from the default CPROPS analysis, while the dot-dashed line shows results from the ECLUMP method.  The vertical dashed line is our adopted completeness limit.  A fit to the shaded histogram is shown, with the corresponding value of $\beta$.}
\label{fig:mhist}
\end{figure}

The differential mass function for GMC-only \cott\ clumps, using the LTE estimate, is shown in Figure~\ref{fig:mhist}.  Following \citet{Williams:97}, we have fitted the distribution with a truncated power law,
\begin{equation}
\frac{dN}{d\ln M} \propto \left(\frac{M}{M_u}\right)^{\beta+1},\qquad
M \leq M_u\;.
\end{equation}
where $\beta$ is the usual power law for the mass distribution ($dN/dM \propto M^{\beta}$).  Guided by the fact that CPROPS can only identify peaks which lie well above the extended background emission from the GMC, we have fitted the distribution only where $\log(M/M_\odot) > 2$, which we adopt as our completeness limit; this corresponds (in the optically thin limit) to a clump with a size twice the beam area and an average column density equivalent to $\bar{I} + 3\sigma$, where $\bar{I}$=40 K \kms\ is the average integrated \cott\ intensity and $\sigma$=0.7 K \kms\ is the RMS noise in the integrated \cott\ map.  In other words, although we are able to detect much lower mass clumps, they would not necessarily be identified as local maxima in the cube given our adopted decomposition algorithm.

The fitted power law, weighted by the inverse variance assuming Poission noise \citep[see][]{Williams:97}, has an index of $\beta = -1.68 \pm 0.05$ for a bin size of 0.2 dex.  Varying the bin size between 0.1 and 0.25 dex yields power laws between $-1.6$ and $-1.75$, suggesting a slightly larger uncertainty in $\beta$ (0.08).  Our derived value of $\beta$ lies within the range of values ($-1.5$ to $-1.9$) that have been previously derived for clumps within molecular clouds using the iterative GAUSSCLUMPS decomposition method, which models the clumps as 3-dimensional Gaussians in position-velocity space \citep[e.g.,][]{Kramer:98,Simon:01}.  It is also consistent with the value of $-1.7 \pm 0.3$ derived by \citep{Mookerjea:04} using CLUMPFIND and GAUSSCLUMPS for the RCW 106 cloud as observed in dust continuum.  Recently, \citet{Reid:06} re-derived the clump mass function in RCW 106 using the data from \citet{Mookerjea:04} and found a double power law with a slope at the high-mass end of $-2.4$; we do not see indication for such a distribution, but further study using denser gas tracers would be valuable.

The dot-dashed histogram shows the clump mass function produced using the ECLUMP method.  Since this method partitions all of the cloud flux among the peaks identified by CPROPS, the resulting clump sizes and masses are significantly larger.  These masses should be viewed with great caution, as the clump sizes are determined largely by the separation between peaks, which in turn was determined largely by our decomposition parameters.  Note that in Paper I, we also found a relatively flat mass function when applying CLUMPFIND to the integrated \cott\ map; this may be due in part to the difficulty of assigning extended emission to clumps.  We believe that the steeper slope found with the CPROPS technique should be considered more robust, since extended emission has been excluded from the clump assignment process.  Nonetheless, the form of the mass function is clearly sensitive to the decompostion method employed, especially for a low-density tracer such as \cott\ which shows both compact and extended emission.

\begin{figure*}
\includegraphics[width=11cm,angle=90,bb=50 170 575 680]{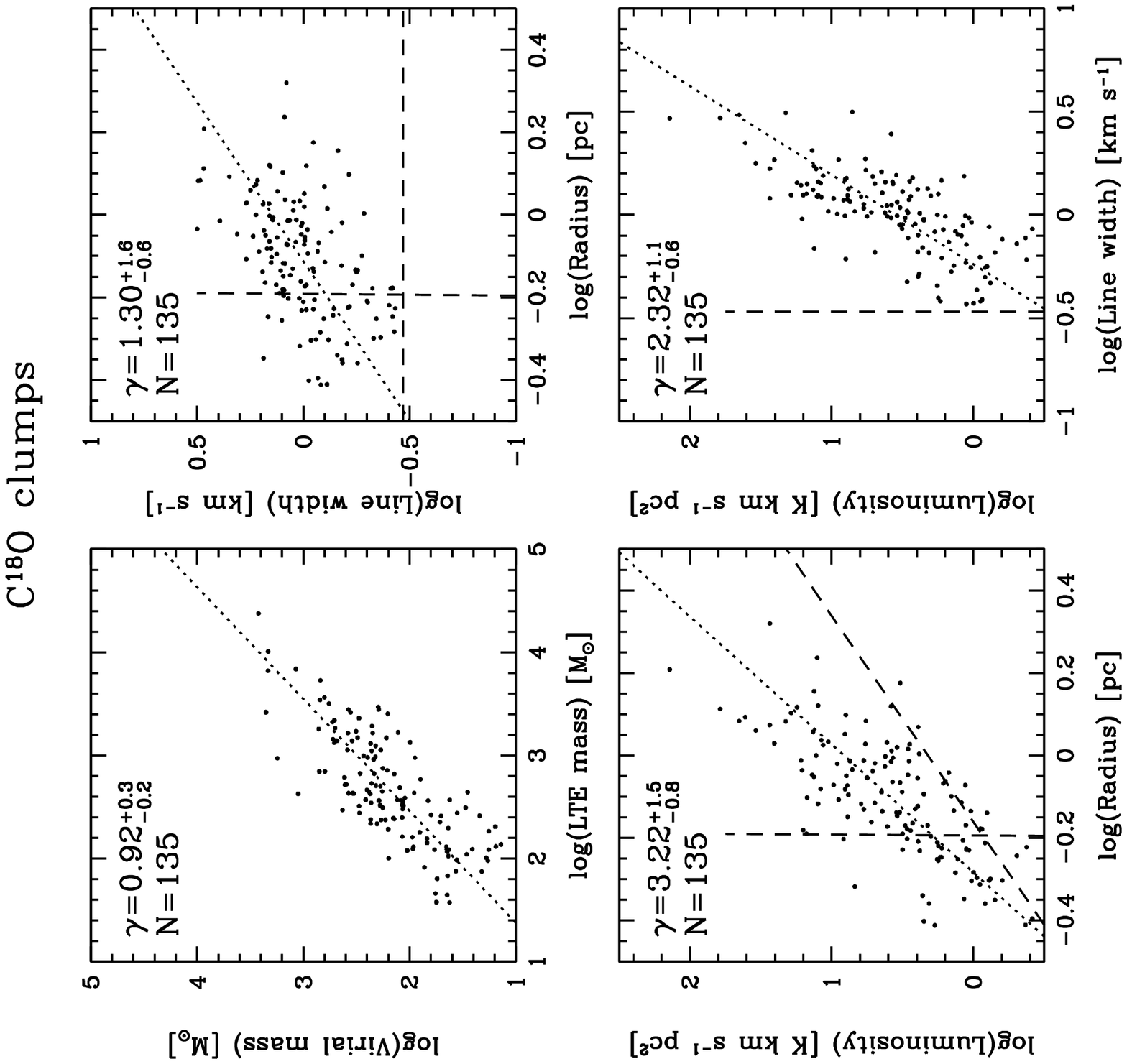}
\bigskip
\includegraphics[width=11cm,angle=90,bb=50 170 575 680]{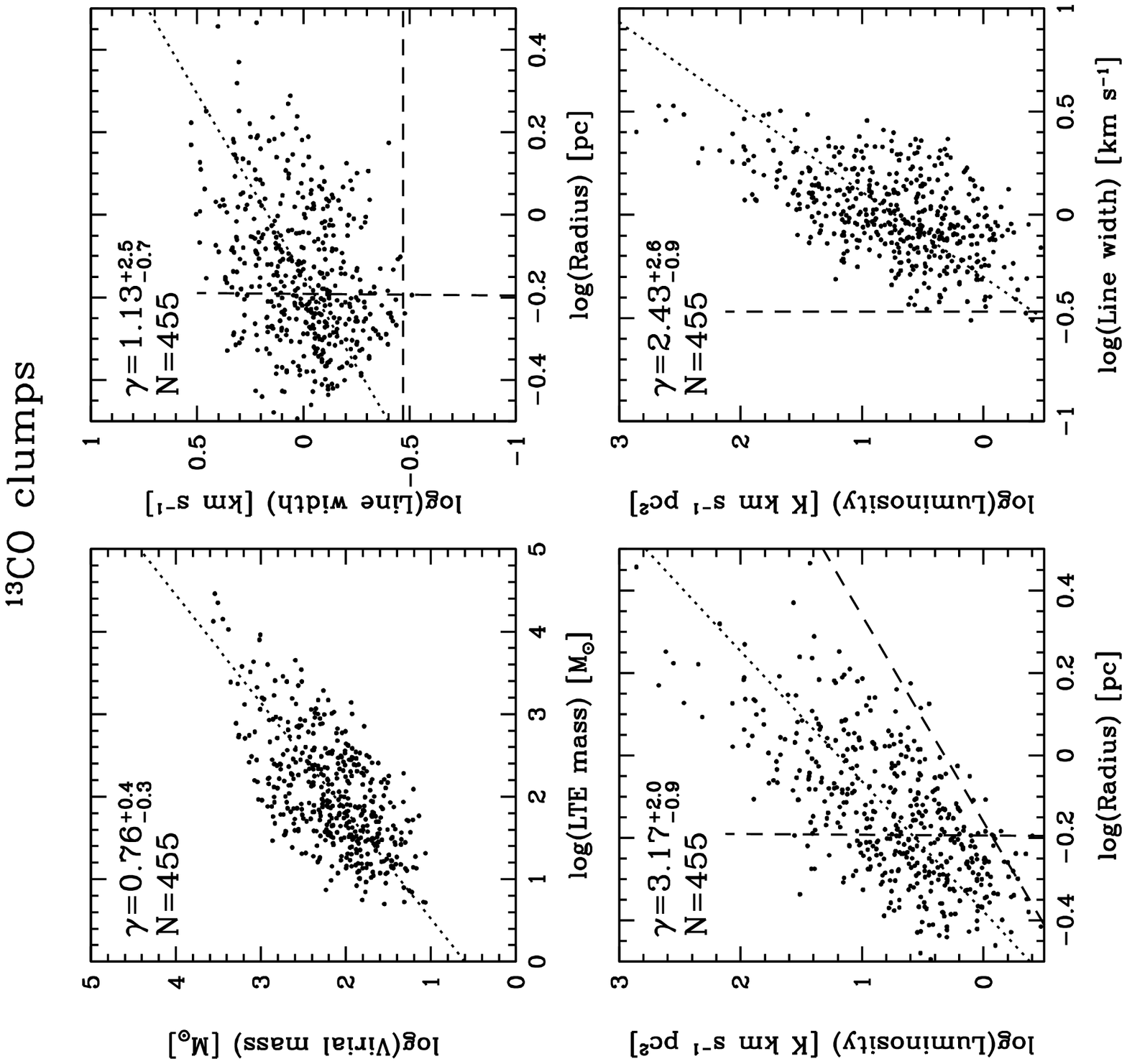}
\caption{({\it Left 4 panels}) Correlation plots of derived clump properties for the GMC-only \cott\ emission.  Dotted lines are OLS bisector fits; the power-law slope is given in the upper left corner.  The range between OLS$(X|Y)$ and OLS$(Y|X)$ is indicated by the quoted errors.  Vertical and horizontal dashed lines represent the instrumental resolution; the skewed dashed line in the luminosity-radius plot represents the sensitivity limit.  ({\it Right 4 panels}) Correlation plots of derived clump properties for the \coet\ emission.}
\label{fig:13cldprps}
\end{figure*}

\begin{figure*}
\includegraphics[width=11cm,angle=90,bb=50 170 575 680]{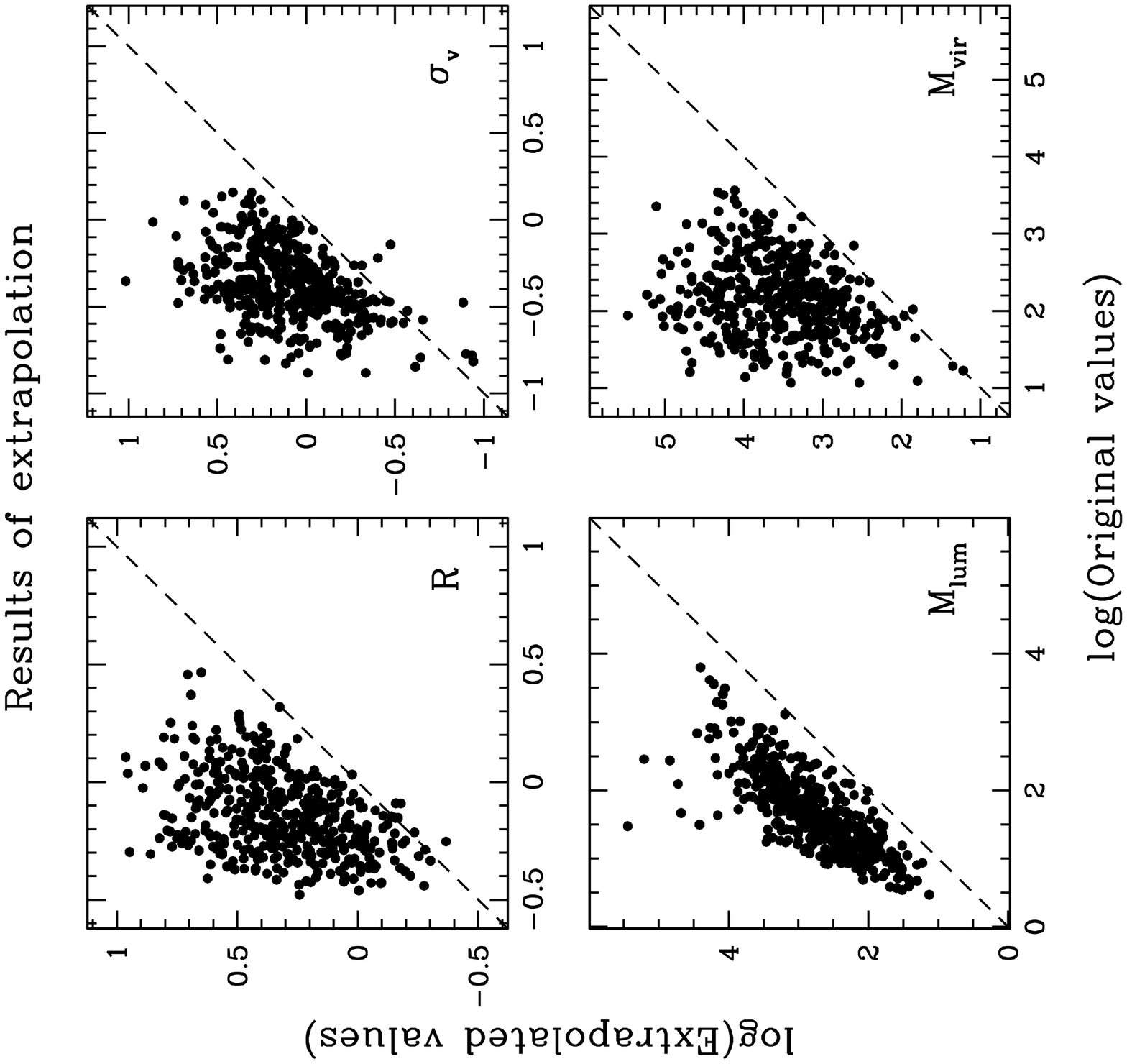}\\
\bigskip
\includegraphics[width=11cm,angle=90,bb=50 170 575 680]{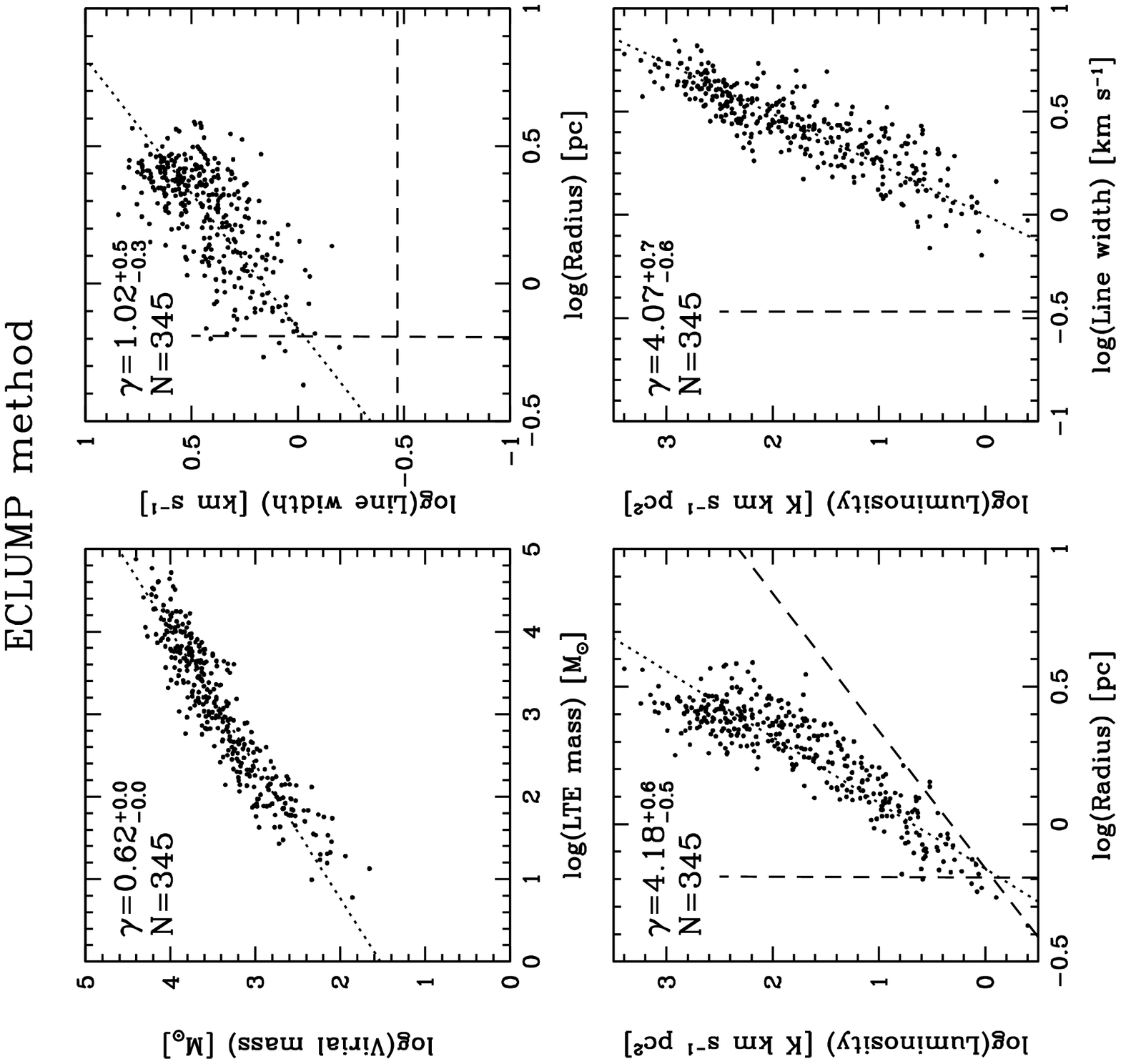}\\
\caption{({\it Left 4 panels}) Correlation plots of derived clump properties for the GMC-only \cott\ emission, using the ECLUMP method.  Dotted and dashed lines are as per Fig.~\ref{fig:13cldprps}.  ({\it Right 4 panels}) Comparison between original values, using the default decomposition method, with the extrapolated values derived by CPROPS.  Extrapolation increases all values significantly, but adds substantial scatter.}
\label{fig:excldprps}
\end{figure*}

\subsection{Size-linewidth and mass-radius correlations}\label{sec:cprops_corr}

In Figure~\ref{fig:13cldprps} we examine, in two sets of four panels each, correlations among clump properties for the \cott\ and \coet\ clumps within the GMC.  Horizontal and vertical dashed lines represent our resolution limits; these are given by $\Delta V$=0.34 \kms\ and $R$=0.64~pc (the latter based on the effective radius for a 45\arcsec\ Gaussian).  For each correlation we compute the slope of the OLS bisector \citep{Feigelson:92}, which treats both variables symmetrically by determining the line which bisects the standard ordinary least-squares fits of $X$ on $Y$, OLS$(X|Y)$, and $Y$ on $X$, OLS$(Y|X)$.  The quoted errors represent the range between the OLS$(X|Y)$ and OLS$(Y|X)$ slopes.  The virial and LTE masses show a clear correlation ({\it upper left panels}) but with considerable scatter, with the \cott\ clumps tending to have larger virial masses compared to LTE masses and the reverse tending to be true for the \coet\ clumps.  We discuss these trends further in \S\ref{sec:avirdisc}.

A weak correlation between linewidth and radius appears to exist ({\it upper right panels}), especially for the \coet\ clumps, although much of the trend is driven by clumps smaller than the nominal resolution limit.  The slope is poorly constrained because of the large scatter in the data points and the limited range of sizes probed.  The fitted slope appears closer to 1 than to 0.5, the latter being the value derived by \citet{Solomon:87} for 273 clouds observed in CO.  Similar studies of correlations for clumps within clouds \citep[e.g.,][]{Kramer:96,Simon:01} also show considerable scatter.  Superposition effects can strongly affect measured clump linewidths, increasing them beyond the actual (3-D) linewidth for a given size scale \citep{Ostriker:01}.

A clear correlation between luminosity and radius is found for both samples ({\it lower left panels}).  While such a correlation is unsurprising, the slope of the correlation ($\gamma \approx 3.2$) appears steeper than would be expected for a constant average column density, which would yield $M \sim M_{\rm LUM} \propto L \propto R^2$.  Including the slight non-linearity between $M_{\rm LTE}$ and $M_{\rm LUM}$ (Fig.~\ref{fig:mlte}) would make the agreement even poorer.  The steepness of the correlation cannot be solely due to the sensitivity limit, which follows an $L \propto R^2$ relation (dashed line).  Excluding clumps based on the resolution and completeness limits does not significantly change the observed slope, although it could then be marginally consistent with $\gamma$=2.  The constant volume (rather than column) density implied by $\gamma \approx 3$ could be related to the critical density needed to excite the CO lines, which is comparable to the average densities of the clumps ($\bar{n}_H \sim 10^{3.3}$ cm$^{-3}$).  Assuming that gas at much higher or lower density is uncommon or poorly traced by CO, one would expect most clumps to have densities near the critical density.

Finally, we note that the luminosity and linewidth show a clear correlation ({\it lower right panels}), although this relation is not independent of the previous two.  

It is worth considering the sensitivity of these relations to the decomposition method employed.  The left side of Figure~\ref{fig:excldprps} shows the correlations for the \cott\ emission using the ECLUMP method.  Here both the virial and LTE masses are considerably larger, presumably a result of the larger clump size.  There is also an even steeper relation of luminosity with radius ($\gamma \sim 4$).  At first glance this result is somewhat surprising, as it implies larger clumps have higher densities.  However, it appears to reflect the tendency for smaller clumps to be found at the edges of the cloud: because of the competitive nature of how flux is assigned to peaks in this method, clumps have a tendency to `grow' in size in the inner parts of the cloud, with the strongest peaks tending to collect the most mass.

One obvious concern raised by Fig.~\ref{fig:13cldprps} is that roughly half (45 per cent) the clump radii fall below the nominal resolution limit set by the Mopra beam.  This is possible because the clumps are identified as intensity peaks, and the size of those peaks above some threshold may be smaller than a beam width.  As noted in \S\ref{sec:propanal}, CPROPS can apply corrections for sensitivity and resolution biases, extrapolating the sizes, linewidths, and fluxes to a zero noise level and then deconvolving the instrumental resolution.  Unfortunately, these corrections do not appear to be reliable for the \cott\ data, as the extrapolated values are poorly correlated with the raw values, and can increase the clump masses by orders of magnitude (Figure~\ref{fig:excldprps}, {\it right}).  Indeed, the sum of the extrapolated clump masses exceeds the total GMC mass by a considerable margin (factor of 1.3--1.6).  Deconvolving the beam without extrapolation excludes many clumps as unresolved, as indicated by the resolution limit in Fig.~\ref{fig:13cldprps}, but the resulting correlations between properties are largely the same, except that the luminosity-radius correlation has a slope closer to the constant column density case ($\gamma \sim 2$) due to a decrease in size of the smallest clumps.

We conclude, in agreement with \citet{Simon:01}, that the classical scaling relations for molecular clouds \citep[e.g.,][]{Solomon:87} are not easily revealed by studies of substructure within clouds, since CO observations probe a limited range of clump size and density.  Subsequent studies making use of molecular lines of higher critical density or dust extinction, both of which can better isolate high-density peaks, should be able to demonstrate these relations more clearly.

\begin{figure}
\includegraphics[width=8.5cm,bb=25 140 580 655]{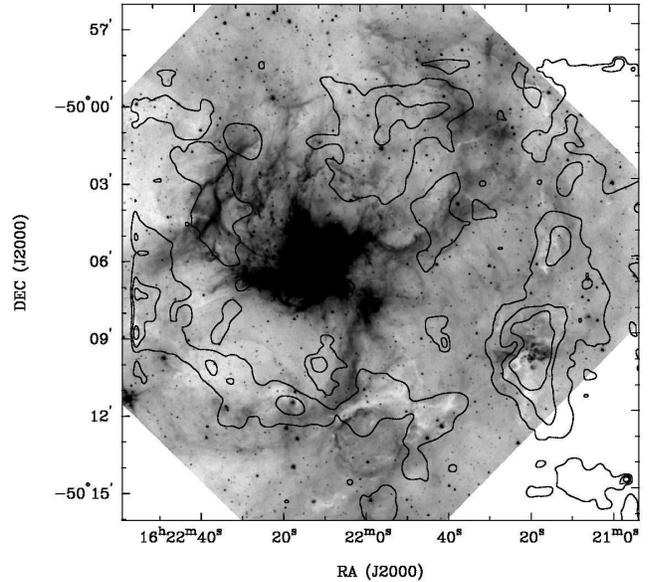}
\caption{Contours of peak \cott\ opacity overlaid on 5.8 $\mu$m GLIMPSE image of the \HII\ region G333.6$-$0.2.  Contour levels represent optical depths of 3, 6, and 9.}
\label{fig:glimpse}
\end{figure}

\subsection{Regions of highest opacity}

As Fig.~\ref{fig:colden} indicates, the regions of highest opacity in the map are not necessarily the regions that have highest \cott\ integrated intensities.  This is not surprising, as high optical depths generally require narrow linewidths as well as significant column density.  The highest optical depths are found in the northern part of the map, in a ring-like structure surrounding the luminous \HII\ region G333.6$-$0.2.  Figure~\ref{fig:glimpse} shows a 5.8$\mu$m image of this region as observed by the Spitzer Space Telescope as part of the GLIMPSE survey \citep{Benjamin:03}.  This \HII\ region has been estimated by \citet{Fujiyoshi:05} to be between $10^5$ and $10^6$ yr old, based on the lack of methanol maser emission \citep{Caswell:97} and the presence of O8V stars.  The ring, although clumpy, may represent a limb-brightened shell driven by stars within the \HII\ region.  At a projected distance of $\sim$9\arcmin\ ($\sim$9 pc) from G333.6$-$0.2, an expansion velocity of $\sim$10 \kms\ would be adequate for the shell to reach its current size within $10^6$ yr.

As shown in Fig.~\ref{fig:glimpse}, several of the high-opacity clouds appear fainter than the background emission in the GLIMPSE image, confirming that they are cold and dense enough to be seen in absorption.  Given recent interest in these `infrared dark clouds' as sites of current or future massive star formation \citep{Simon:06,Rathborne:06}, a more detailed study of this complex would be worthwhile.

\section{Discussion}

\begin{figure}
\includegraphics[width=8.5cm,bb=33 353 553 679]{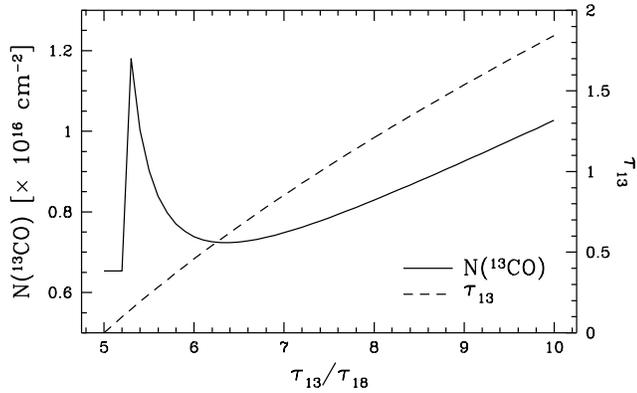}
\caption{Dependence of column density and opacity on the adopted abundance ratio $\tau_{13}/\tau_{18}$.  Both curves are drawn assuming \tmb(\cott)=5~K and \tmb(\coet)=1~K.  The discontinuous jump in $N$ at left is due to the assumption of \tex=20~K in the optically thin limit.}
\label{fig:abvar}
\end{figure}

\begin{figure*}
\includegraphics[width=5.8cm,bb=30 155 554 680]{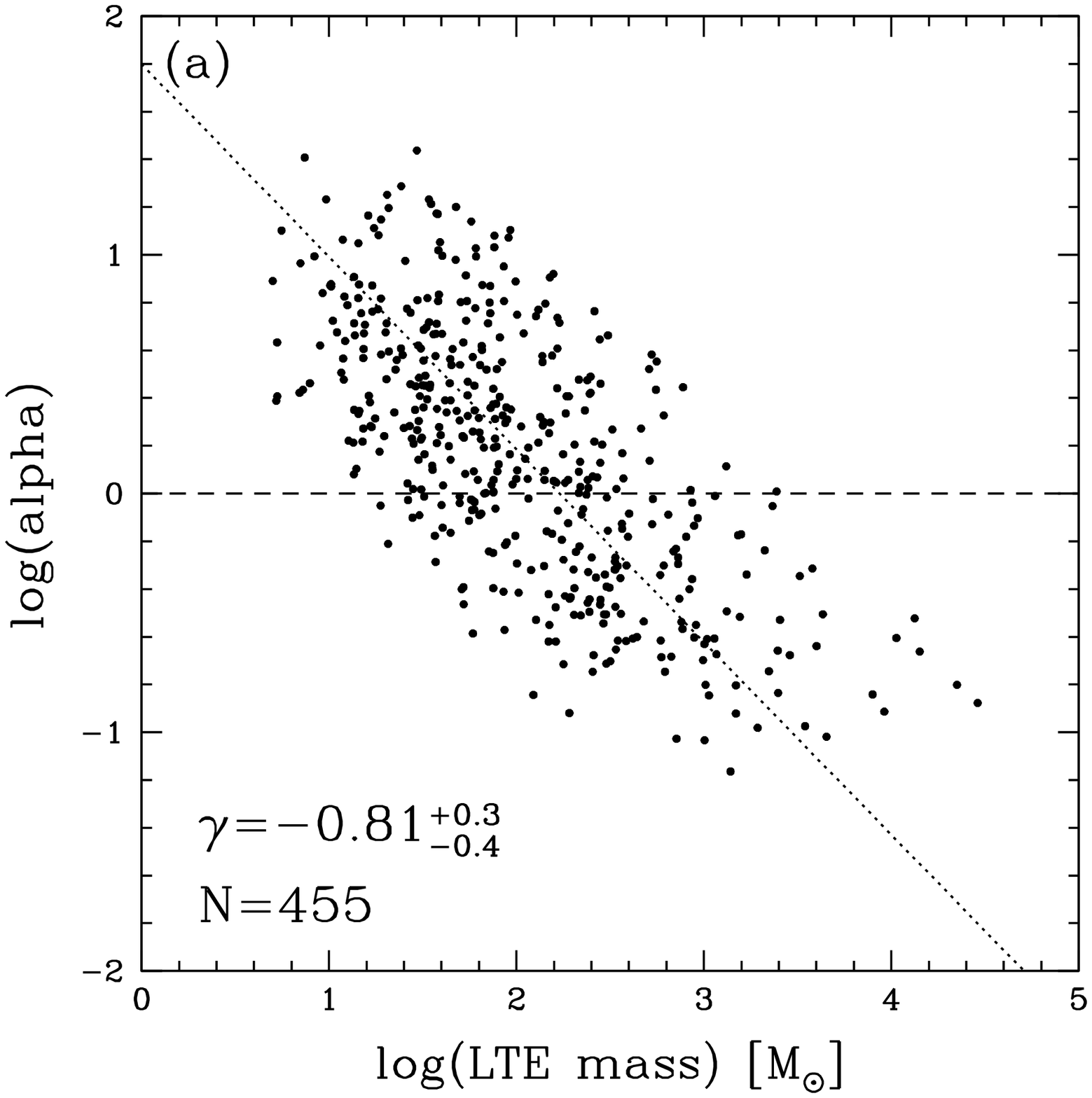}\hfill
\includegraphics[width=5.8cm,bb=30 155 554 680]{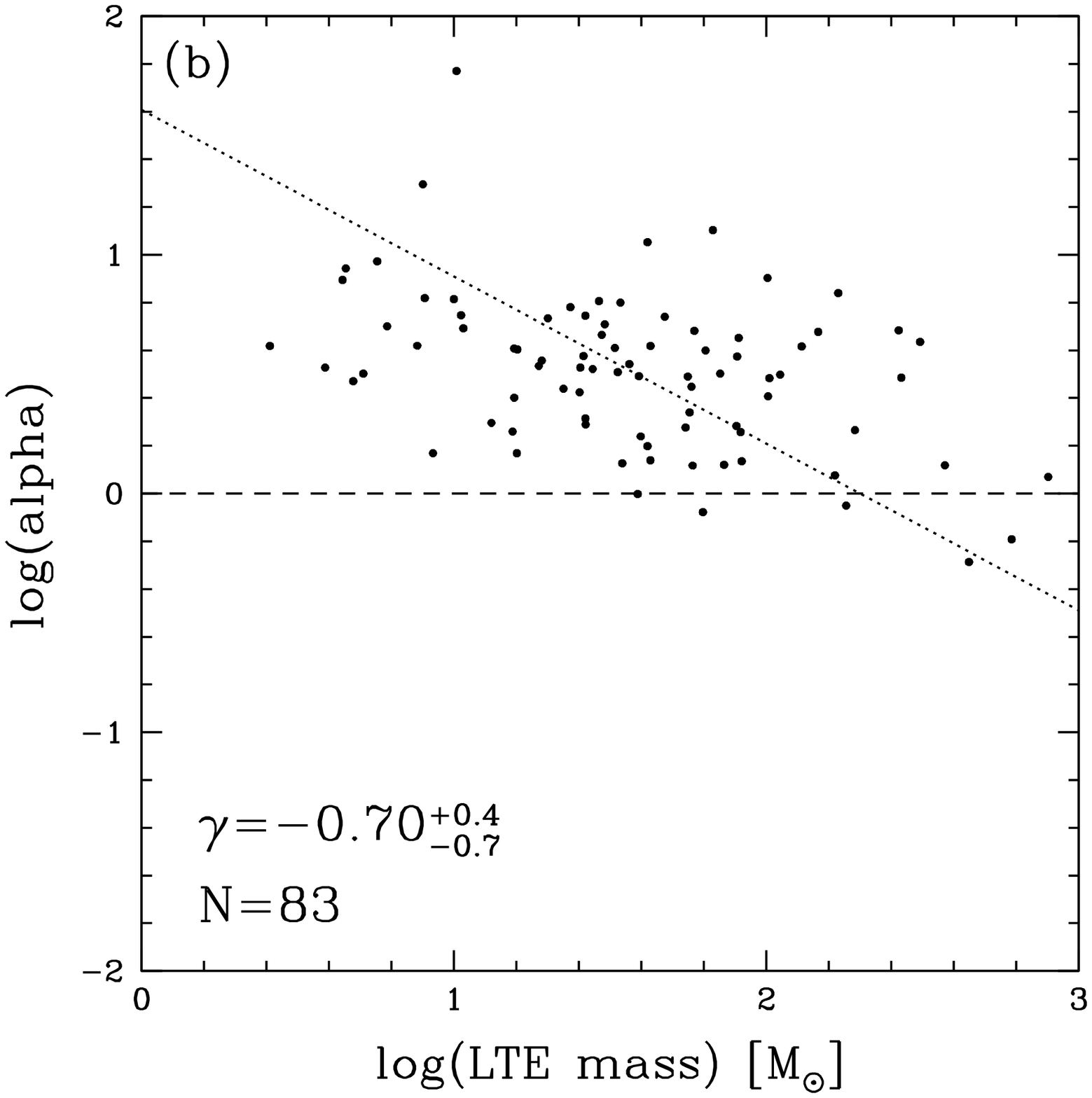}\hfill
\includegraphics[width=5.8cm,bb=30 155 554 680]{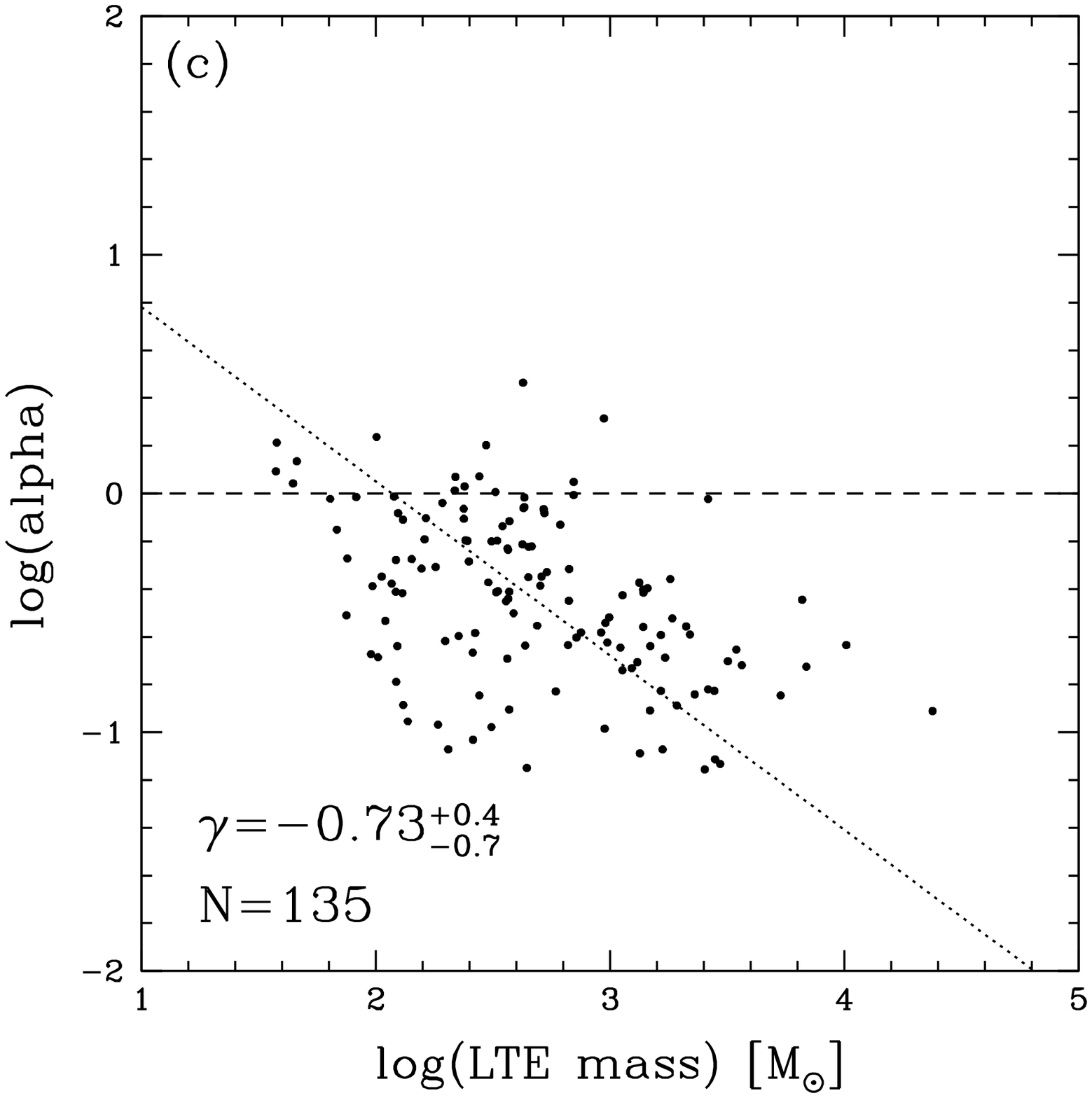}
\caption{Virial parameter defined in Eq.~\ref{eqn:alpha} plotted against LTE mass for (a) \cott\ emission from the GMC; (b) non-GMC \cott\ emission; (c) \coet\ emission.  Dashed lines are OLS bisector fits.  The number of clumps and slope of each fit are indicated on each plot.  Note the scale of the horizontal (mass) axis differs among the panels.}
\label{fig:avir}
\end{figure*}

\subsection{Uncertainties in column density}\label{sec:uncdisc}

One limitation of this analysis is the lack of an independent constraint on the kinetic (and thus excitation) temperature at the resolution of the Mopra data.  Pointed \cotw\ observations by one of us (BM) suggest peak \tas(\cotw) values of up to 20--30 K, which translates to 40--60 K (\tmb).  Such brightnesses are roughly a factor of 2 larger than the corresponding peak \cotw\ brightness temperatures seen in the NANTEN data.  In the optically thin approximation, the column density for a thermalized population increases by roughly a factor of $\sim$5 for \tex\ increasing from 10--100~K.  Thus we underestimate the column density by at most a factor of $\sim$2 in the warmer regions.  With a new spectrometer that is capable of observing \cotw, \cott, and \coet\ simultaneously, future Mopra observations should be able to better constrain the variation in \tex\ across clouds.  Even with matched resolution observations, however, there would remain some ambiguity about whether the same material was being traced by \cotw, given its high optical depth.

An even greater limitation in warmer regions of the cloud is the use of the LTE approximation, especially in populating the unobserved higher rotational levels.  Direct measurements of higher $J$ transitions are still required to test the LTE assumption.  In nearby globules, \citet{Harjunpaa:04} find that the LTE approximation overestimates column densities by 10--30 per cent, due to subthermal excitation of the higher rotational levels.  In massive GMCs like RCW106, where overall temperatures are likely to be higher, the total column density becomes even more sensitive to the column densities in excited rotational states, and thus the uncertainty in extrapolating from $N(J=0)$ is likely to become more significant.  On the other hand, volume densities are also likely to be higher, which may help to thermalize the higher $J$ transitions.  Future observations with submillimetre telescopes such as NANTEN2 will be important in determining whether a non-LTE approach is required.

Finally, uncertainties and possible variations in the \cott/\coet\ abundance ratio will affect the inferred opacity and column density values.  As noted in \S\ref{sec:tau}, about 12 per cent of the voxels have \cott/\coet\ intensity ratios exceeding our adopted abundance ratio of 7.4.  Unusually high abundances of \cott\ relative to other CO isotopomers have been noted in the outskirts of molecular clouds \citep{Dickman:79,Langer:80}, and have been considered to result from chemical fractionation, in which isotopic exchange reactions ($^{13}$C$^+$ + \cotw\ $\leftrightarrow$ $^{12}$C$^+$ + \cott) favor \cott\ at low temperatures \citep{Watson:76,Langer:77}.  For the isotopic exchange to be effective, the medium must be partially irradiated by UV photons (yielding a significant abundance of ionised carbon) whilst still being cold enough to suppress the reverse exchange reaction.  Both conditions could be fulfilled at the edges of a cloud.  Figure~\ref{fig:abvar} shows how the LTE column density (cf.\ Eqs.\ \ref{eqn:tau18}, \ref{eqn:tmbdef}, \ref{eqn:ntot}) varies for \cott/\coet\ abundance ratios ranging from 5 to 10, assuming \tmb(\cott)=5~K, \tmb(\coet)=1~K, and a default \tex=20~K (in the optically thin limit).  Whereas $\tau_{13}$ is very sensitive to the adopted abundance ratio, the column density $N$(\cott) is much less so, as a result of the anti-correlation between $\tau$ and \tex\ for a given observed brightness temperature.

We conclude that errors in the column density of a factor of $\ga$2 could quite plausibly result from uncertainties in excitation temperature, isotopic abundance ratio, and the LTE approximation, with the greatest uncertainties in warmer regions or cloud edges, where the column density may be underestimated.

\subsection{The PDF of the column density}

The probability distribution function (PDF) of the volume density in a turbulent medium is well-described by a log-normal distribution, provided that the medium is approximately isothermal \citep{Vazquez:94,Padoan:97,Passot:98}.  Even if the turbulence is not strictly isothermal, the PDF should remain close to log-normal, developing a quasi-power-law tail \citep{Passot:98}.  However, volume densities are difficult to infer observationally, requiring observations of several $J$ transitions to be interpreted within the context of a radiative transfer model.  In order to facilitate comparison with observations, \citet{Ostriker:01} and \citet{Vazquez:01} have investigated the PDF of the {\it column} density in molecular cloud models.  For the isothermal case, the column density PDF remains log-normal, at least until the line-of-sight depth becomes much larger than the autocorrelation function of the density field \citep{Vazquez:01}.  Our observation of a log-normal column density PDF lends support to the view that this form of the PDF is a generic consequence of interstellar turbulence.

Recently, \citet{Lombardi:06} have investigated the column density PDF in the Pipe nebula, based on near-infrared (near-IR) extinction towards background stars.  They also find an approximately log-normal PDF, although there are additional peaks in the distribution, some of which may be related to additional clouds along the line of sight.  Comparing their results with previous \cotw\ mapping, they find that \cotw\ fails to trace very low column densities (due presumably to UV dissociation) and very high column densities (due presumably to saturation or depletion).  \citet{Ridge:06} also find that near-IR extinction yields a much closer match to a log-normal PDF than \cotw\ and \cott\ line intensities in Ophiuchus and Perseus.  Our ability to recover a log-normal distribution from obsevations of CO isotopomers is likely a reflection of three effects: (1) we correct for optical depth, whereas previous studies have tended to ignore opacity; (2) our observations are confined to largely molecular regions where \cott\ is likely tracing the bulk of the gas mass; and (3) the GMC contains many active \HII\ regions, where depletion of CO onto grains is probably less important than in nearby dark clouds.  Thus, the approach we outline here may be useful for studying the column density PDF in clouds near the Galactic plane where extinction-based methods face greater difficulties.

\subsection{Dynamical state of the clumps}\label{sec:avirdisc}

A commonly used indicator of the dynamical state of clumps is the virial parameter \citep{Bertoldi:92}, 
\begin{equation}
\alpha = \frac{5\sigma_v^2R}{GM} \approx \frac{M_{\rm VIR}}{M_{\rm LTE}}\;,
\label{eqn:alpha}
\end{equation}
which measures the ratio of kinetic to gravitational energies.  For $\alpha \gg 1$, the clump must be confined by external pressure in order to survive as a long-lived entity, whereas $\alpha \sim 1$ indicates that the clump is close to self-gravitating.  \citet{Bertoldi:92} found that for $\alpha>1$ (pressure-confined clumps), the virial parameter correlates with cloud mass as approximately $\alpha \propto M^{-2/3}$.

Figure~\ref{fig:avir} shows plots of $\alpha$ vs.\ $M_{\rm LTE}$ for the GMC-only \cott\ clumps, the non-GMC \cott\ clumps, and the \coet\ clumps.  All show a trend of decreasing $\alpha$ with $M$ that is roughly consistent with the slope of $-2/3$ predicted by \citet{Bertoldi:92}.  Somewhat unexpectedly, however, the more massive clumps within the GMC, particularly the \coet\ clumps, show values of $\alpha \ll 1$, and thus seem unable to support themselves against gravity.  By contrast, the non-GMC clumps, which we identify with foreground or background Galactic gas, appear to have lower masses and to be either pressure supported or marginally self-gravitating.

The low values of $\alpha$, down to $\sim$0.1, are also puzzling given that studies of clumps within other clouds typically find very few clumps with $\alpha \la 1$ \citep{Kramer:96,Heyer:01,Simon:01}.  One possibility is that we are overestimating the gas mass by assuming thermalisation of all $J$ levels.  However, Table~\ref{tbl:mass} suggests that this can at most explain a factor of 2 discrepancy, and it would be counteracted by our likely underestimate of the excitation temperature (\S\ref{sec:uncdisc}).  A second possibility is that we underestimate the virial mass because we measure the size and linewidth of each clump only near the emission peak, rather than extrapolating to infinite sensitivity.  However, such extrapolation increases $M_{\rm LUM}$ and $M_{\rm LTE}$ as well, and thus appears to have little net effect on $\alpha$.  A third possibility is that we overestimate the distance, since $\alpha \propto D^{-1}$.  Since we adopt the near kinematic distance, however, it seems unlikely that the molecular complex can be much nearer.  While none of these effects seem able to explain the low $\alpha$ values alone, a combination of them might be a viable solution.

If values of $\alpha \ll 1$ are not an artifact of our analysis technique, then there must be additional terms in the virial equation that are counteracting gravity.  For instance, there may be a significant degree of magnetic support for the clumps, as suggested by \citet{Bot:07} for clouds in the Small Magellanic Cloud with similarly low values of $\alpha$.  In this scenario, one would still need to explain why the $\alpha$-$M$ relation shows a smooth variation from the pressure-confined regime ($\alpha>1$) to the magnetically supported regime ($\alpha<1$), and how a strong degree of magnetic support is maintained in the most massive clumps, where ambiopolar diffusion should be most effective because of lower ionisation fractions in more shielded regions of the cloud.  In addition, as noted below (\S\ref{sec:padisc}), the axis ratios of the clumps are not consistent with an oblate population, as would be expected in the simplest magnetically supported models.

Alternatively, \citet{Dib:07} have criticized the use of the virial parameter to assess the dynamical state of clumps and cores, because of its neglect of the surface terms in the virial equation, which represent the exchange of energy between a region and its environment.  A useful illustration is the notion of turbulent pressure, which is generally assumed to provide support against gravity, yet turbulent compression might plausibly act to accelerate collapse.  The net effect of turbulent motions depends sensitively on the surface kinetic energy term, which is difficult to constrain observationally.  The simulations of \citet{Dib:07} show $\alpha \propto M^{-0.6}$ even for $\alpha < 1$.  Yet many of the simulated clumps with $\alpha < 1$ are not truly bound, i.e.\ they have positive total energy, and are thus likely to be transient.  This is especially true of the lower density regions that would be probed by CO.  In future work, we plan to investigate the $\alpha$-$M$ relation for clumps identified in tracers of higher gas densities to test this possibility.

\begin{figure}
\includegraphics[width=8.5cm]{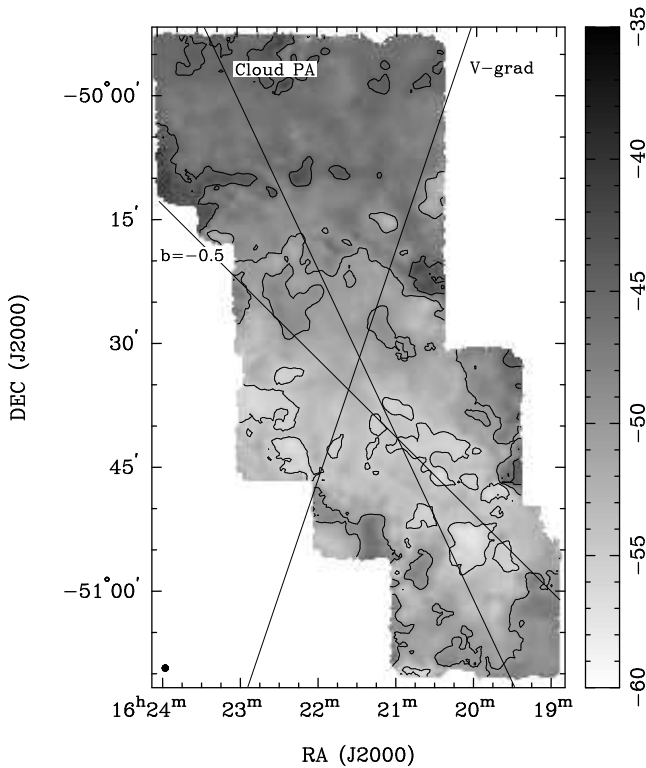}
\caption{Intensity-weighted mean velocity image of the GMC-only \cott\ emission, with approximate position angles of the cloud and the velocity gradient overlaid.  A line of Galactic latitude $b$=$-0.5$\degr\ is also shown.  Contours range from $-60$ to $-40$ \kms\ in increments of 5 \kms.}
\label{fig:vgrad}
\end{figure}

\subsection{GMC velocity gradient and clump elongations}\label{sec:padisc}

\citet{Bains:06} (Paper I) had noted a velocity gradient across the cloud of $\sim$0.2 \kms\ arcmin$^{-1}$ by inspection of position-velocity slices.  We measured the gradient to be 0.13 \kms\ arcmin$^{-1}$ by fitting a plane to an image of the intensity-weighted first moment along the velocity axis.  Since 1\arcmin$\approx$1~pc at our adopted distance, this is consistent with the typical velocity gradients, projected on the plane of the sky, of 0.15 \kms\ pc$^{-1}$ for molecular clouds in the Galactic ring \citep{Koda:06}.  As noted by \citet{Koda:06}, such gradients are substantially larger than the gradient expected from Galactic rotation (0.04 \kms\ pc$^{-1}$).  Indeed, large scale velocity gradients, inconsistent with Galactic rotation, are commonly revealed by principal component analysis of CO data cubes \citep{Brunt:03}.  These gradients may be an indication of turbulence driven on large scales, as microturbulent models have difficulty accounting for such gradients \citep{Brunt:03}.

The observed velocity gradient (Figure~\ref{fig:vgrad}) has a position angle of $-19$\degr\ (counterclockwise relative to north) and appears therefore somewhat misaligned with the elongation of the cloud (at P.A.$\sim$25\degr); it makes an angle of 64\degr\ relative to the Galactic plane, whereas the misalignment of the cloud itself relative to the plane is only $\sim$20\degr.  Since the cloud has not been mapped completely to its edges, the orientation of the velocity gradient is not strongly constrained.  Nonetheless, an alignment of the gradient with the Galactic plane appears unlikely.  This is consistent with the analysis of \citet{Koda:06}, who found that molecular clouds in the inner Galaxy are preferentially elongated along the Galactic plane, but that their spin axes have random orientations.  They argue that this elongation cannot be due to rotation and must result from internal velocity anisotropy, and thus favor a driving source for turbulence related to Galactic rotation.  However, it is unclear to what extent the tendency for clouds to be elongated along the plane is related to internal motions rather than the initial conditions for cloud formation in a vertically stratified gas layer.  

Focusing on the clumps {\it within} the RCW106 cloud, with typical sizes of $\sim$1~pc and line widths of $\sim$1~\kms, the large-scale velocity gradient appears to play only a minor role in accounting for the observed velocity dispersions.  Whether the clumps themselves have significant velocity gradients is unclear; since they are only slightly resolved in most cases, reliable fitting is difficult and measurement biases would need to be carefully quantified.  Of the four brightest \coet\ clumps, three show gradients of $\ga$0.5 \kms\ arcmin$^{-1}$ when fitting a plane to the first moment image, with position angles of $-31$\degr, 36\degr, and 118\degr.  We suspect that clump velocity gradients will show only a weak correspondence with the large-scale velocity gradient, just as the {\it morphological} position angles show little alignment with the cloud as a whole or the Galactic plane (Fig.~\ref{fig:pahist}).  Thus, clumps and their properties appear to be defined primarily by turbulence within the cloud.

As pointed out by \citet{Ryden:96}, mean apparent axis ratios of 0.5--0.6, as found by the present work and previous work \citep[e.g.,][]{Myers:91} are fully consistent with a population of prolate objects, but only marginally consistent with a population of oblate objects.  Therefore static, magnetically supported models, which yield oblate equilibria unless the magnetic field has a significant toroidal component \citep{Fiege:00}, are probably not applicable for the clumps we observe.  Further support for this conclusion comes from work by \citet{Gammie:03}, who applied a clump-finding algorithm to 3-D simulations of magnetohydrodynamic turbulence.  Their distribution of apparent axis ratios peaks around 0.5, and only 10 per cent of their clumps are oblate.  However, our observations lack the spatial resolution to study the elongation of star-forming {\it cores}, where the importance of magnetic support is most controversial.

\section{Conclusions}

In this paper we have presented \coet\ mapping of the central part of the GMC associated with RCW 106, which had been previously mapped at similar resolution in \cott\ (Paper I).  The \cott\ and \coet\ data have been used to estimate the optical depth and excitation temperature as a function of position and velocity, and thereby obtain estimates of the column density assuming LTE.  In a separate analysis, we have decomposed both \cott\ and \coet\ datacubes into clumps using an automated decomposition algorithm and estimated the clump masses using the LTE column density cube.  The principal results of this study are as follows:

\begin{enumerate}
\item Correcting for opacity does not have a dramatic effect on the overall cloud mass, since regions of high opacity are restricted to compact regions.  Depending on the adopted excitation temperature, assuming negligible opacity can even lead to an overestimate of the cloud mass.

\item Correcting the \cott\ emission for opacity does broaden the distribution of column densities, which otherwise exhibits a relatively rapid falloff at the high end.  The resulting column density PDF is well-approximated as a log-normal function, consistent with what is commonly found in numerical simulations of turbulence.

\item The regions of highest opacity are found in the northern part of the cloud, in a ring-like structure surrounding the luminous \HII\ region G333.6$-$0.2.  We speculate that this ring is a limb-brightened shell, $\sim$10 pc in radius, driven by the expansion of the \HII\ region.  The shell material is cold enough to appear in absorption in mid-infrared {\it Spitzer} images.

\item The distribution of clump masses can be approximated as a power law, $dN/dM \propto M^{-1.7}$, consistent with previous studies of clumps within molecular clouds.  While this result appears sensitive to the decomposition method employed, we argue that the adopted method is less prone to assigning unrelated extended emission to the largest clumps.

\item Positive correlations between clump size, linewidth, and luminosity are found, albeit with considerable scatter.  The luminosity-radius correlation appears closer to $L \propto R^3$ than $L \propto R^2$, suggesting most clumps have a volume density close to the critical density needed to excite CO emission.  The strength and interpretation of such correlations should be revisited in future work with maps tracing higher density gas.

\item The virial parameter $\alpha$, which measures the ratio of kinetic to gravitational energies, is inversely correlated with the LTE clump mass.  The most massive clumps in the GMC show $\alpha \ll 1$, indicating that another source of support against gravity is required, possibly magnetic fields.  Alternatively, $\alpha$ may fail to capture the true dynamical state of the clumps, due to the neglect of time-varying surface terms in the virial theorem.

\item The clumps show mean apparent axis ratios of $\sim$0.5, with no preferential alignment along the Galactic plane.  On the other hand, the GMC as a whole does appear elongated along the plane, although it exhibits a large-scale velocity gradient with a different orientation.

\end{enumerate}

Future studies in this series will probe the turbulent properties of the cloud using the power spectrum and related techniques, and incorporate observations of higher-density tracers, as well as the young stellar population, to address directly the relationship between molecular cloud properties and star formation.

\section*{Acknowledgments}

We thank the ATNF staff, particularly J\"urgen Ott, Michael Kesteven, Bob Sault, and Mark Calabretta, for help with Mopra characterisation and data reduction.  We also appreciate assistance from Erik Rosolowsky and Annie Hughes with using CPROPS and IDL.  We thank the anonymous referree for helpful suggestions which led to an improved manuscript.  T.W. was supported by an ARC-CSIRO Linkage Postdoctoral Fellowship and subsequently by the University of Illinois.

\bibliographystyle{mn}
\bibliography{atnf}

\end{document}